\documentclass[fleqn,usenatbib]{mnras}
\UseRawInputEncoding
\usepackage{xcolor}
\usepackage{newtxtext,newtxmath}
 
\usepackage[T1]{fontenc}

\DeclareRobustCommand{\VAN}[3]{#2}
\let\VANthebibliography\thebibliography
\def\thebibliography{\DeclareRobustCommand{\VAN}[3]{##3}\VANthebibliography}

\usepackage{graphicx}	
\usepackage{amssymb}
\usepackage{amsmath}	
\usepackage{threeparttable}

\definecolor{address}{rgb}{0.36,0.54,0.66}
\definecolor{red}{rgb}{0.8,0.,0.}

\newcommand{\HII}{\mathrm{H}\,\textsc{\large{ii}}}
\newcommand{\HI}{\mathrm{H}\,\textsc{\large{i}}}
\newcommand{\HIs}{\mathrm{H}\,\textsc{\textmd{i}}}
\newcommand{\HIb}{\mathbf{H}\,\textsc{\large{i}}}

\newcommand{\Hmol}{\mathrm{H_2}}

\newcommand{\kms}{\mathrm{km\,s}^{-1}}

\newcommand{\cc}{\mathrm{cm}^{-3}}
\newcommand{\K}{\mathrm{K}}
\newcommand{\eagle}{\textsc{\large eagle}{ }}
\newcommand{\eagleNS}{\textsc{\large eagle}}
\graphicspath{{Plots/}}

\title[H\,\textsc{\Large i} line profiles of \textsc{eagle} galaxies]{Drivers of asymmetry in synthetic H\,\textsc{\huge i} emission-line profiles of galaxies in the \textsc{eagle} simulation}
\author[A. Manuwal et al.]{Aditya Manuwal,$^{1}$\thanks{E-mail: aditya.manuwal@icrar.org}
Aaron D. Ludlow,$^{1}$
Adam R. H. Stevens,$^{1}$
Ruby J. Wright$^{1}$ and 
Aaron S. G. Robotham$^{1}$
\\
$^{1}$International Centre for Radio Astronomy Research, The University of Western Australia, 35 Stirling Highway, Crawley, WA 6009, Australia\\
}
\date{Accepted 2021 November 29. Revised 2021 November 22; Received: 2021 September 23}
\pubyear{2021}

\begin{document}
\label{firstpage}
\pagerange{\pageref{firstpage}--\pageref{lastpage}}
\maketitle

\begin{abstract}
  We study the shapes of spatially integrated $\HI$ emission-line profiles of galaxies in the \eagle simulation using
  three separate measures of the profile's asymmetry. We show that the subset of \eagle galaxies whose gas fractions
  and stellar masses are consistent with those in the xGASS survey also have similar $\HI$ line asymmetries. Central galaxies with
  symmetric $\HI$ line profiles typically correspond to rotationally supported $\HI$ and stellar disks, but those with asymmetric line profiles
  may or may not correspond to dispersion-dominated systems. Galaxies with symmetric $\HI$ emission lines are, on average, more
  gas rich than those with asymmetric lines, and also exhibit systematic differences in their specific star formation rates, suggesting that turbulence generated by stellar or AGN feedback may be one factor contributing to $\HI$ line asymmetry.
  The line asymmetry also correlates strongly with the dynamical state of a galaxy's
  host dark matter halo: older, more relaxed haloes host more-symmetric galaxies than
  those hosted by unrelaxed ones. At fixed halo mass, asymmetric centrals tend to be surrounded by a larger number of massive subhaloes than their symmetric counterparts, and also experience higher rates of gas accretion and outflow. At fixed stellar mass, central galaxies have, on average, more symmetric $\HI$ emission lines than satellites; for the latter, ram pressure and tidal stripping are significant sources of asymmetry.
\end{abstract}




\section{Introduction}
Neutral atomic hydrogen ($\HI$) is the dominant component of cold galactic gas and can be observed through the 21-cm
emission resulting from the spin-flip transition of the atom's sole ground-state electron. The global or unresolved $\HI$ line
profile of a galaxy [i.e. the $\HI$ flux density as a function of line-of-sight (LOS) velocity] carries combined information about the
spatial distribution \citep[e.g][]{Huchtmeier1988} and dynamics \citep[e.g][]{Stilp2013} of its $\HI$ gas. The 21-cm emission is
optically thin and largely unaffected by dust extinction, which makes it an ideal tracer of $\HI$ in galaxies at low redshifts
($z<0.4$). As a result, it has become a widely used tool in extra-galactic astronomy for exploring the link between atomic hydrogen and galaxy evolution.

A stable, rotating gas disk, oriented approximately edge-on, will exhibit a canonical double-horned $\HI$ line profile.
Internal disturbances (e.g. non-circular motions, warps, etc.) in the disk may manifest as asymmetries in such profiles, but ones that should disappear after a few
rotation periods due to the disk's differential rotation \citep{Baldwin1980}. Asymmetric line profiles can also arise due
to galaxy-galaxy interactions or ram pressure in cluster environments \citep{Scott2018}. Compared to isolated
central galaxies, satellite galaxies are more prone to tidal and ram-pressure stripping and, as a
result, typically have $\HI$ velocity profiles that are more asymmetric \citep[e.g.][]{Reynolds2020,Watts2020}. Nevertheless, $\HI$
surveys reveal a high incidence ($\gtrsim 50$ per cent) of asymmetric line profiles among massive spiral galaxies
\citep{Richter1994,Matthews1998,Watts2020}, even seemingly isolated ones \citep{Haynes1998,Espada2011,Portas2011}; similar
trends are also observed for dwarf galaxies \citep{Swaters2002}. This suggests that $\HI$ line asymmetries are
not ephemeral, but are the result of prolonged processes that require explanation.

A reasonable approach to understanding their origins of asymmetric $\HI$ emission lines is through self-consistent modelling of galaxies and
their dark matter haloes within their large-scale environments. This can be achieved using state-of-the-art cosmological,
hydrodynamical simulations. Such simulations have already been used to study the topology of 21-cm emission from the
pre-reionization era \citep{Kuhlen2006} and the 21-cm signal from the Epoch of Reionization \citep{Baek2009}, the distribution
of $\HI$ in high-redshift galaxies \citep{Duffy2012} and how it relates to environment  at lower redshifts \citep{Marasco2016,Stevens2019}.

On smaller scales, simulations have been used to study the link between $\HI$ and molecular clouds in the outer Galaxy
\citep{Douglas2010}, the effect of outflows on the $\HI$ content of galaxies \citep{Dave2013}, the observable properties
and physical conditions of the Galactic interstellar $\HI$ \citep{Kim2014,Murray2017,Fukui2018}, $\HI$ surface density
profiles \citep{Bahe2016,Stevens2019origin}, and the origin of spiral arms in $\HI$ far beyond the optical radius of a galaxy
\citep{Khoperskov2016}. Other studies used the $\HI$ content of simulated galaxies to study gas kinematics \citep{El2018},
dwarf galaxy rotation curves \citep{Oman2019}, and damped Lyman-alpha absorbers \citep{eagledla}. Mock $\HI$ line profiles
have also been used to reconcile the observed $\HI$ velocity function with those obtained from cosmological simulations
\citep[e.g.][]{Maccio2016,Brooks2017,Chauhan2019}.

Simulations have also been used to study characteristics of the unresolved $\HI$ line profiles of galaxies. \citet{Watts2020b},
for example, used the IllustrisTNG simulations \citep[][hereafter TNG100]{Nelson2019} to investigate the relationship between line asymmetry and
environment, using halo mass as a proxy for the latter. They found that satellite galaxies are on average less symmetric than
centrals, but the trend appears to be dominated by satellites hosted by haloes with masses $\gtrsim 10^{13}$~M$_\odot$. 

\citet{Deg2020} simulated equilibrium galaxy models and investigated the impact of velocity resolution and signal-to-noise ratio ($S/N$) on three
separate asymmetry measures, and determined
which measure correlates best with visually classified asymmetry.
They noted that, for a particular galaxy -- even an asymmetric one -- there is often an orientation for which its
line profile appears symmetric, making it difficult to
draw conclusions from observations about whether a particular galaxy harbours a symmetric distribution of $\HI$ gas. Based on the analysis of
$\HI$ lines obtained from the Westerbork $\HI$ survey of Irregular and SPiral galaxies (WHISP), \citet{van2001} found the
``channel-by-channel'' asymmetry (obtained by comparing velocity channel pairs equally offset to the low- and high-velocity side of a
central or systemic velocity) to correlate best with the asymmetry in the flux map, and to better reflect non-axisymmetric galactic features.

In this paper, we use the \eagle simulation to explore possible factors contributing to the asymmetry of global $\HI$ emission 
profiles. We begin by calculating the $\HI$ content of each galaxy and its associated $\HI$ line profile. We then quantify the global
line profile asymmetries for well-resolved galaxies in \eagle and investigate differences in the galaxies and haloes 
corresponding to symmetric and asymmetric systems. We also investigate the impact of observational effects on inferred asymmetries,
including the LOS and distance to a galaxy, as well as the effective 
velocity resolution and $S/N$ ratio of its line profile.
We compare the unresolved line profiles of \eagle galaxies to those obtained from the extended \textit{GALEX} Arecibo SDSS Survey
\citep[hereafter xGASS;][]{Catinella2018}, finding good agreement between the two data sets.

The paper is organized as follows. We provide details about the xGASS survey in Section~\ref{observe}; relevant information about the
\eagle simulations is provided in Section~\ref{sim}. Our primary analysis techniques are described in Section~\ref{analyse}, including
a detailed description of how we model the $\HI$ content of \eagle galaxies (Section~\ref{himod}) and their associated unresolved $\HI$ emission-line 
profiles (Sections \ref{hiprof} and \ref{measure}). Our main results are presented in Section~\ref{results}, which includes an assessment
of how numerical resolution and observational effects impact the inferred shapes of line profiles (Sections~\ref{resolution} and \ref{obseff},
respectively), a comparison of the line profiles of central galaxies in \eagle and xGASS (Section~\ref{reconcile}), and a detailed look at
the physical processes that may give rise strongly asymmetric line profiles for both central and satellite galaxies (Sections~\ref{centrals}
and Section~\ref{satellites}, respectively). We end with a summary of our main results in Section~\ref{summary}. 
  
\section{Observational data: the {\small x}GASS survey}\label{observe}

xGASS \citep{Catinella2018} is a survey of the 21-cm emission from atomic hydrogen in $\approx 1200$ galaxies in the local Universe
($0.01 < z < 0.05$); it is an extension of the GASS survey \citep{Catinella2010} to lower galaxy stellar mass (GASS probed the stellar mass range
$10^{10}\,{\rm M}_\odot < M_\star < 10^{11.5}\,{\rm M}_\odot$ whereas xGASS evenly sampled galaxies across the range
$10^9\,{\rm M}_\odot < M_\star < 10^{11.5}\,{\rm M}_\odot$). Each observation corresponds to a single beam pointing, yielding an
unresolved global emission line profile for each target galaxy. Each line has a velocity resolution of
$\Delta v = 1.4~\kms$ at 1370~MHz, and is Hanning-smoothed to a lower \textit{effective} resolution
$\Delta v_\mathrm{sm}$ (which varies from $4$ to $63\,\kms$, depending on the $S/N$ ratio) in order to account
for the Gibbs ringing phenomenon \citep{Van1989} and to aid in the identification of the line profile's peaks and edges.

xGASS galaxies are located at the
intersection of the SDSS DR7 \citep{SDSS}, \textit{GALEX} Medium Imaging Survey (MIS; \citealt{GALEX}) and ALFALFA \citep{ALFALFA} footprints.
They were observed until detected, or until a gas-to-stellar mass fraction of a few per cent could be ensured as an upper limit; specifically, the limiting $\HI$ masses
are $M_{\HIs}$/$M_\star > 0.02$ for $M_\star > 10^{9.7}$~M$_\odot$, and $M_{\HIs} = 10^{8}$~M$_\odot$ for
$M_\star \leq 10^{9.7}$~M$_\odot$. Note that all xGASS galaxies used in the analysis that follows were identified as central galaxies
in the SDSS DR7 Group B\,\textsc{\large ii} catalogue \citep{Yang2007,Janowiecki2017}.
As suggested by \citet{Watts2020}, we also exclude all galaxies in xGASS that were detected with a signal-to-noise
ratio $S/N<7$, as these may incur biased asymmetry measurements. 

\section{Numerical simulations}\label{sim}

\subsection{The \eagle simulations}
The \eagle project \citep{Schaye2015,Crain2015} consists of a suite of cosmological, hydrodynamical simulations run with a modified version of
the \textsc{\large{gadget}}-3 code \citep{Springel2005}. We here describe aspects of the simulations and their sub-grid physics
that are relevant to our study; additional information can be found in \citet{Schaye2015} and \citet{Crain2015}, among other papers.

All of our results are based on the intermediate-resolution simulation of a periodic cube of comoving side length $L=100$ Mpc
in which the linear density field was sampled with ${\rm N}=1504^3$ particles of both dark matter and baryons (this run is
referred to as Ref-L100N1504 in \citealt{Schaye2015}). The simulation assumed a flat $\Lambda$CDM cosmology with cosmological
parameters consistent with the \citet{Planck2014} results:  $h = H_0/(100$~km s$^{-1}$ Mpc$^{-1})$ $=0.6777$
is the (dimensionless) Hubble constant; $\Omega_\text{m,0} = 0.307$, $\Omega_\text{b,0} = 0.04825$ and $\Omega_{\Lambda,0} = 0.693$ are the present-day
cosmological density parameters of matter, baryons and dark energy, respectively (expressed in units of the critical density of
the Universe, i.e. $\rho_{\rm crit}=3H_0^2/8\,\pi\,G$, where $G$ is the gravitational constant); $\sigma_8 = 0.8288$ is the linear rms density
fluctuation in 8-Mpc spheres; $n_s=0.9661$ is the power spectral index of primordial density fluctuations; and $Y=0.248$ is the
primordial helium abundance. With this set-up the masses of dark matter and (primordial) baryonic particles are
$m_\text{dm} = 9.70 \times 10^6$~M$_\odot$ and $m_\text{g} = 1.81\times 10^6$~M$_\odot$, respectively. The gravitational
softening length was held fixed at a value of $\epsilon=2.66$ comoving kpc until $z=2.8$, but remained fixed in physical coordinates
thereafter (i.e. $\epsilon=0.7$ kpc for $z\leq 2.8$).

Photoionization and radiative cooling rates for gas particles were computed using the scheme of \citet{Wiersma2009}, assuming a
\citet{Haardt2001} extra-galactic photo-ionizing background. The Jeans scale in the warm inter-stellar medium (ISM) is marginally resolved,
which precludes detailed modelling of the cold gas phase. In order to limit artificial fragmentation of gas particles, a temperature floor
$T(\rho_{g})$ corresponding to the equation-of-state $P\propto \rho_g^{4/3}$ was imposed, which was normalized so that $T_{\rm eos}=8000~\K$ at
$n_\text{H} = 0.1~\cc$ ($n_\text{H}$ is the total hydrogen number density). 

Star formation is implemented by stochastically converting gas particles into stellar particles \citep[see][]{Schaye2008}, 
assuming a metallicity-dependent density threshold \citep{Schaye2004}. Each stellar particle represents a stellar population 
with a \citet{Chabrier2003} initial mass function. The transport of metals from stellar particles into the ISM was modelled
following \citet{Wiersma2009b}.

Feedback from supernovae and active galactic nuclei (AGN) were calibrated so that the simulation reproduced the observed galaxy stellar mass
function  \citep[GSMF; e.g.][]{Li2009,Baldry2012} and the galaxy size--mass relation at $z=0.1$ \citep[e.g.][]{Shen2003}.

\subsection{Identification of dark matter haloes and galaxies}\label{halos}

Dark matter haloes were identified using a friends-of-friends (FoF) algorithm \citep{Davis1985} employing a linking
length $b = 0.2$ times the Lagrangian mean inter-particle separation. The star, black hole and gas particles are associated with the FoF
group corresponding to their nearest dark matter particle, provided that particle belongs to one. The self-bound substructures
(or subhaloes) in these groups were identified using \textsc{\large subfind} \citep{Springel2001,Dolag2009}. We refer to the most massive subhalo in
each group as the central subhalo, which hosts the central galaxy; the remaining substructures are ``satellite'' subhaloes, which host satellite galaxies.

For each halo and subhalo -- and their associated central and satellite galaxies -- \textsc{\large subfind} identifies the particle at $\vec{r}_\mathrm{p}$ with the minimum gravitational potential, which we identify as its centre. For FoF halos (and their central galaxies) we define the virial radius $r_{200}$ as that of the sphere, centred on that particle,
in which the enclosed density is $200\times \rho_\text{crit}$; the corresponding virial mass is $M_{200}$ (note that
we include all particles of each type when calculating $M_{200}$ and not just those bound to the central subhalo). The mass of
a satellite subhalo, $m_{\rm sub}$, is defined as the total mass that \textsc{\large subfind} deems gravitationally bound to it (again including
all particle types). Note that for central and satellite subhaloes, \textsc{\large subfind}
also calculates other quantities of interest, for example the location and magnitude of their peak circular velocities, $R_{\rm max}$
and $V_{\rm max}$, respectively. 

The stellar mass of a galaxy (central or satellite) is defined as the integrated mass of all bound stellar particles
within a 30 (physical) kpc spherical aperture centred on its host halo, which is similar to that enclosed by a 2D Petrosian aperture \citep{Schaye2015}. The total gas mass is defined by the sum of all gravitationally bound gas particles enclosed by a 
70 kpc spherical aperture (i.e. for centrals we exclude gas particles bound to nearby satellites,
and for satellites we exclude gas particles bound to the background host halo when calculating their gas masses). This aperture roughly corresponds to the full width at half-maximum beam size of the Arecibo L-band Feed Array
\citep[ALFA;][]{Giovanelli2005} at the median redshift of xGASS (i.e. $z\approx 0.03$).

Due to the nature of
the FoF algorithm -- which sometimes artificially links multiple nearby haloes -- some ``satellite'' subhaloes are located beyond
$r_{200}$ of their host halo, often at unexpectedly remote distances; a few, for example, even exceed $2\times r_{200}$.
Because these systems may not have experienced the same environmental effects as ``genuine'' satellite galaxies associated with the halo \citep[see e.g.][]{Bakels2021}, we exclude them from 
the analysis that follows. This removes $\approx 30$ per cent of \textsc{\large subfind}
satellites from our {\it final} sample, and does not unduly bias our results.

\subsection{Constructing halo merger trees}

For all FoF haloes (and their central galaxies) merger trees were constructed following the approach of \citet{Qu2017}. The method forward-tracks
particles within each halo across consecutive simulation snapshots to determine their descendants. Any given halo and its complete list of
descendants forms a branch of a merger tree, which begins at the first snapshot in which it was identified and extends until it has
either merged with a more massive halo or to $z=0$, whichever is sooner. The merger tree of a given halo constitutes the complete
set of branches of its surviving subhaloes. We use these merger trees to construct mass accretion histories for the dark
matter haloes hosting central galaxies, which are defined using virial mass of their main progenitors, i.e. $M_{200}(z)$. 

\section{Analysis}\label{analyse}

\subsection{Modelling the neutral atomic hydrogen content of \eagle galaxies}\label{himod}

The \eagle simulations do not explicitly model the total or neutral atomic hydrogen content of baryonic particles, which would
require a more detailed treatment of radiative transport than is feasible at \textsc{\large eagle}'s resolution. We instead follow an
approximate scheme (employed by \citealt{Crain2017}) and partition each particle's mass into atomic and 
molecular hydrogen, $\Hmol$, using a modified version of open-source modules described in \citet{Stevens2019} (note that \eagle
self-consistently models the particle mass fractions in the form of helium and nine metal species, in addition to hydrogen). 

We start by partitioning the hydrogen component of each gas particle into its neutral (i.e. atomic plus molecular) and ionized
($\HII$) states. We use the empirical prescription of \citet{Rahmati2013}, which
was calibrated using \textsc{\large traphic} \citep{traphic} radiative transfer simulations and predicts both the collisional and
photoionization of hydrogen. The photoionization rate accounts for the combined effects of the metagalactic ultra-violet background
radiation (UVB, with associated photoionization rate $\Gamma_{\rm UVB}$), self-shielding and diffuse radiative recombination. The
prescription uses the total hydrogen number density, self-shielding threshold, and $\Gamma_\textrm{UVB}$ to calculate
the {\em effective} photoionization rate, $\Gamma_\textrm{Phot}$ \citep[see equation~(A1) in][]{Rahmati2013}. Along with the temperature and
hydrogen number density of gas particles, $\Gamma_\textrm{Phot}$ is used to compute the neutral gas fraction
\citep[equation~(A8) in][]{Rahmati2013}, and includes the effects of collisional ionization. We model the $z=0$ metagalactic UVB background
using the prescription of \citet{Haardt2012}, which is lower by about a factor of 3 than that predicted by the \citet{Haardt2001} model adopted
for \eagleNS; the impact on the predicted $\HI$ masses of galaxies is, however, negligible \citep[see][for details]{Crain2017}.
Since \eagle lacks the resolution required to model the multi-phase ISM we do not consider the local ionizing radiation field within the galaxies.
We instead adopt a fixed temperature for star-forming gas particles of $T = 10^4~\K$ in order to mimic the warm, diffuse ISM around young stellar
populations when computing the ionization states \citep{Crain2017}.

\begin{figure}
  \begin{center}
    \includegraphics[width=1\columnwidth,trim={0cm 0cm 0cm 0cm},clip=true]{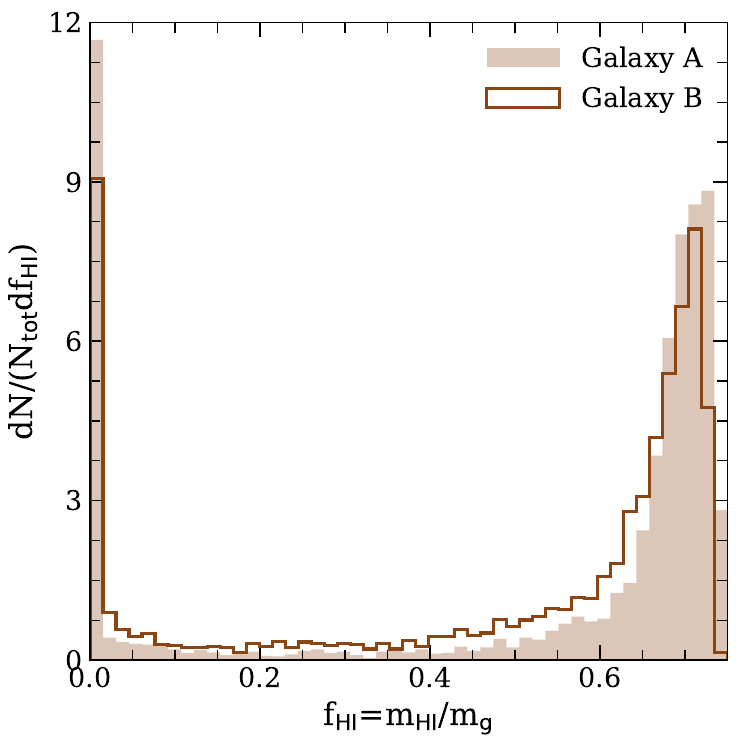}
    \caption{The distribution of $\HI$ fractions, $f_{\HIs}\equiv m_{\HIs}/m_{\rm g}$, for gas particles associated with two
      galaxies in the \eagle simulation (note that these are the same galaxies used below for Figs.~\ref{illust} and~\ref{pois}).
      The bimodal distribution
      of $f_{\HIs}$ is typical of all galaxies in our sample and comprises two distinct particle populations that are either
      $\HI$-rich (defined as $f_{\HIs}\geq 0.5$) or $\HI$-poor ($f_{\HIs}<0.5$), where the latter contain mostly ionized gas. Note that over 90 per cent of the
      $\HI$ mass of galaxies in our sample is typically contributed by $\HI$-rich particles.}
    \label{fhihist}
  \end{center}
\end{figure}

The next step is to partition the neutral hydrogen into its atomic and molecular components. To do so, we tested the (empirical and
analytic) models of \citet[][hereafter BR06]{BR06}, \citet{L08}, \citet{GK11}, \citet{K13} and \citet{GD14}
(the first two models are based on scaling relations between the molecular fraction and the ISM mid-plane gas pressure; the latter three,
described in detail in \citealt{Diemer2018} and \citealt{Stevens2019}, involve detailed modelling of the formation and destruction of molecular hydrogen).
Despite the differences in their implementation, we found that all five prescriptions result in galaxy $\HI$ masses that are comparable across the halo mass range considered in our study.
BR06, in particular, has been used 
in previous studies of $\HI$ content of \textsc{eagle} galaxies, and 
shown to yield reasonable masses and sizes for their $\HI$ disks \citep[e.g.][]{Bahe2016,Crain2017}. For that reason, 
we will present results obtained using that prescription. 

The BR06 model
relates the molecular fraction $f_{\rm mol}$ of gas particles to the ISM mid-plane gas pressure $P$ using the empirical scaling relation
\begin{equation}
  f_\textrm{mol} \equiv \frac{\Sigma_\Hmol}{\Sigma_{\HIs}} = \left(\frac{P}{P_0}\right)^\alpha,
  \label{eq:fmol}
\end{equation}
where $\Sigma_{\Hmol}$ and $\Sigma_{\HIs}$ are the surface densities of ${\rm H_2}$ and $\HI$, respectively, and  $P=n_\mathrm{H} \,k_\mathrm{B} \,T$ is the pressure of gas particles at temperature $T$ ($k_\mathrm{B}$ is the Boltzmann constant), $P_0/k_\mathrm{B}=4.3\times10^4~\cc$~K and $\alpha=0.92$. We
use this relation to approximate the molecular-to-atomic hydrogen mass ratio in each gas particle.

Applying equation~(\ref{eq:fmol}) to gas particles in \eagle galaxies results in a strongly bimodal distribution of their atomic 
hydrogen fractions, $f_{\HIs}\equiv m_{\HIs}/m_{\rm g}$ ($m_{\HIs}$ is the particle's $\HI$ mass). 
We show this in Fig.~\ref{fhihist}, where we plot the
distribution of $f_{\HIs}$ for all gas particles that lie within a 70 kpc spherical aperture centred on two \eagle galaxies (these galaxies were selected to span the extremes of the $\HI$ line asymmetries of \eagle galaxies and are used for illustration purposes in several sections that follow). We find that
over $90$ per cent of their total $\HI$ mass is contributed by gas particles with $f_{\HIs} \geq 0.5$, which is typical of the entire
population of galaxies used in our analysis. In what follows, we will refer to gas particles with $f_{\HIs} \geq 0.5$ as ``$\HI$-rich'' particles. 

\subsection{Modelling the unresolved $\HIb$ line profiles  of \eagle galaxies}\label{hiprof}

\subsubsection{$\HI$ line profiles without instrumental noise}\label{himod_noiseless}

To construct the $\HI$ line profile of an \eagle galaxy we first select the gas particles that
a) \textsc{\large subfind} deems gravitationally bound to it,\footnote{Note that including {\em all} gas particles within a 70 kpc aperture (rather than only those bound to the galaxy) typically alters their asymmetries by $\lesssim 20$ per cent. Importantly, the impact of substructure on asymmetry is random in nature and is independent of the asymmetries inferred using only the gravitationally-bound mass.} and b) lie within 70 kpc of its
centre of potential. We then project their velocity vectors along the LOS to obtain their LOS velocity distribution (the orientation of the LOS will be described in Section~\ref{losasym}). To account for thermal broadening, 
we assume that each particle has an intrinsic (three-dimensional) velocity
dispersion equal to $\sigma_\mathrm{T}=\sqrt{k_\mathrm{B}\,T/m_\mathrm{p}}$ ($m_\mathrm{p}$ is the proton mass). We use this to divide each particle's $\HI$ mass into discrete LOS velocity bins, allowing individual particles to overlap multiple velocity bins if necessary. 

The resulting $\HI$ line profiles correspond to the distribution of the galaxy's $\HI$ mass in bins of LOS velocity, i.e. $\delta M_{\HIs}(v)$; 
their shapes are therefore sensitive to viewing angle. As we discuss below, a thin, rotationally supported $\HI$ disk viewed edge-on
will exhibit a broad double-horned profile. When viewed face on it instead exhibits a much narrower Gaussian-like profile.
We elaborate on projection effects and how they impact line profile asymmetries in more detail in Section~\ref{losasym}.

\subsubsection{Creating mock observations of $\HI$ line profiles}\label{himod_noisy}

\begin{figure*}
  \begin{center}
    \includegraphics[width=1.7\columnwidth,trim={0cm 0cm 0cm 0cm},clip=true]{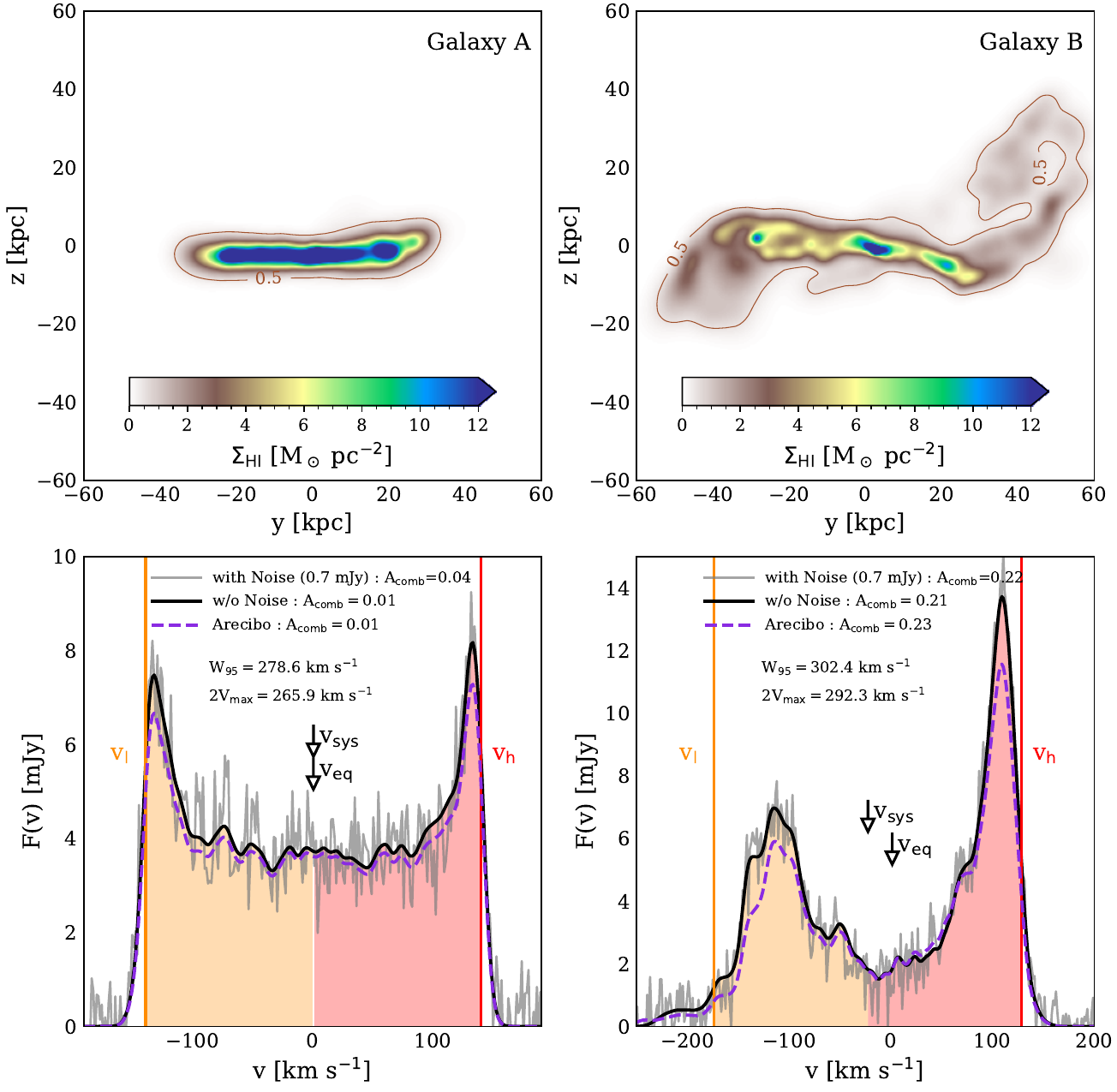}
    \caption{Examples of the $\HI$ surface density (upper panels) and corresponding $\HI$ emission-line profiles (lower panels)
      for two galaxies in the \eagle
      simulation. The contours in the upper panels indicate a surface density of $0.5~{\rm M}_\odot~{\rm pc}^{-2}$.
      Galaxy A (left panels) has a smooth rotationally supported $\HI$ disk,
      whereas Galaxy B (right panels) is visibly disturbed. The solid black lines in the lower panels
      correspond to the $\HI$ mass profiles binned by LOS velocity and 
      converted to $\HI$ flux using equation~(\ref{profile}) (we assume $D=129$~Mpc, $z=0.027$, $\Delta v_{\rm sm}=1.4~\kms$ and $N(v)=0$; where $D$ is the median distance of galaxies in xGASS). The thin grey line shows the same profiles after adding
      instrumental noise corresponding to the median noise level of xGASS observations (i.e. $\sigma_{\rm rms}=0.7$~mJy). The purple dashed lines show the profiles obtained assuming a single-dish
      Arecibo observation (no noise is added in this case, i.e. $N(z)=0$). The vertical lines and arrows
      marks various characteristic velocities that are used to estimate the profile's asymmetry: 
      $v_\mathrm{l}$ (orange vertical line) and $v_\mathrm{h}$ (red vertical line) are, respectively, the low- and high-velocity edges
      the profile; $v_\mathrm{sys}$ marks the systemic velocity (i.e. the mid-point velocity between $v_\mathrm{l}$
      and $v_\mathrm{h}$); $v_\mathrm{eq}$ is the velocity with respect to which the lopsidedness
      asymmetry, $A_{\rm l}$, is minimized (see equation~\ref{eq:Al} and Section~\ref{lop} for details). The orange and pink shaded regions are,
      respectively, the low- and high-velocity sides of the profile that are used to determine various asymmetries described in Section~\ref{measure}. In both bottom
      panels, we list the profile width ($W_{95}$), the maximum circular velocity of galaxy's dark matter halo (more specifically, $2\, V_\mathrm{max}$), and 
      several values of the line profile's combined asymmetry parameter, $A_\mathrm{comb}$ (see Section~\ref{measure} and equation~\ref{eq:Acomb}
      for details).}
    \label{illust}
  \end{center}
\end{figure*}

In order to meaningfully compare our simulated line profiles to observed ones -- and to test the impact of
various observational effects such as instrumental noise, distance to the galaxy and 
effective velocity resolution -- we convert the $\delta M_{\HIs}(v)$
profiles to the corresponding $\HI$ {\em flux} line profiles using \citep[see][for details]{Catinella2010}
\begin{equation}
\frac{F(v)}{\mathrm{Jy}} = \frac{1+z}{2.356\times10^5}\frac{\delta M_{\HIs}(v)/\mathrm{M}_\odot}{(D/\mathrm{Mpc})^2\,(\Delta v/\mathrm{km~s}^{-1})} + \frac{N(v)}{\mathrm{Jy}}.
\label{profile}
\end{equation}
Here $F(v)$ is the $\HI$ flux density in the velocity bin centred on $v$, $\Delta v$ is the bin's
width, $D$ is the luminosity distance at redshift $z$, and $N(v)$ is an instrumental noise term.
We adopt a velocity resolution compatible with Arecibo observations,\footnote{\citet{Deg2020} showed that estimates of 
  asymmetry are unreliable when line profiles are sampled with fewer than $\approx 20$ velocity channels, regardless of velocity bin width
  and $S/N$. Our adopted bin width of $\Delta v=1.4~\kms$ and the characteristic velocities of galaxies in our sample ensure that
  all the profiles used in this study are sampled by $>20$ channels, with most having $> 100$. For the xGASS sample, we only include
  galaxies whose raw-spectrum emission lines are resolved with $>20$ velocity channels.}
i.e. $\Delta v=1.4~\kms$, and we assume that the instrumental noise in all velocity bins is Gaussian distributed (the rms noise is fixed for each mock line profile).
Our spectra are smoothed\footnote{The smoothing is carried out using a one-dimensional box kernel and is implemented using the
  \textsc{B\large{ox}}\textsc{1DK}\textsc{\large{ernel}} function in \textsc{\large astropy} \citep{Astropy}}
over a fixed number of velocity channels, $N_\mathrm{sm}=\Delta v_\mathrm{sm}/\Delta v$, in order
to achieve a desired {\em effective} velocity resolution $\Delta v_\mathrm{sm}$.
This is often done in observational studies to facilitate the identification of the edges and peaks of line profiles.
In practice, we add noise prior to smoothing using input noise level $\sigma_{\rm{rms},i}=\sigma_\mathrm{rms}\sqrt{N_\mathrm{sm}}$,
which ensures that the {\em smoothed} profile has the desired level of noise, i.e. $\sigma_{\rm rms}$.
Note that noiseless flux profiles (i.e. those for which $\sigma_{\rm rms}=0$) are equivalent to velocity-binned
$\HI$ mass profiles, i.e. $\delta M_{\HIs}(v)$, up to a normalization constant.

The black lines in the lower panels of Fig.~\ref{illust} compare the (noiseless) $\HI$ line profiles of one symmetric and one asymmetric
galaxy (left- and right-hand panels, respectively). Both galaxies are viewed edge-on, i.e. perpendicular to the total angular momentum
vector of their $\HI$ gas disks, and we assume $D=129$~Mpc (which is the median galaxy distance in xGASS)
and $\Delta v_{\rm sm}=1.4~\kms$. The thin grey lines show the effect of instrumental noise, assuming $\sigma_{\rm rms}=0.7\,{\rm mJy}$.
For comparison, the corresponding upper panels show the $\HI$ surface density maps constructed using \textsc{\large py-sphviewer} \citep{pysphviewer}. Galaxy A (left) has a thin, rotationally supported $\HI$ disk, whereas Galaxy B's disk (right) is visibly disturbed. As expected for edge-on disks, the
resulting $\HI$ lines resemble the canonical double-horned profile, albeit an asymmetric one in the case of Galaxy B (lower panels).

The Arecibo dish has an approximately Gaussian-shaped beam with a sensitivity that decreases from the centroid outwards, and may
therefore underestimate the true $\HI$ flux originating from parts of a galaxy that are
farther away from the beam centre. This alters the shapes of $\HI$ line profiles. For a disky galaxy, the LOS velocity increases with
increasing (projected) distance from the galaxy centre. If its centre is coincident with the beam centre, this results in dampened
flux in velocity bins closer to the edges of the beam. The effect is shown explicitly in the lower panels of Fig.~\ref{illust} using
purple dashed lines, which show the line profiles obtained when explicitly modelling the Arecibo beam. Although these profiles differ slightly from the ones that do not model Arecibo's beam, the impact on the inferred asymmetry is generally small, less than about 20 per cent in most cases. We therefore neglect beam effects in what follows.

\subsection{Quantifying the asymmetries of unresolved $\HI$ line profiles}\label{measure}

This section provides a detailed description of how we quantify the line widths and asymmetries of our $\HI$ line profiles.
Although this can be achieved in a variety of ways \citep[e.g.][]{Peterson1974,Tifft1988,Haynes1998,Deg2020,Reynolds2020}, we focus on 
statistics that capture asymmetry on different scales and then combine them, for convenience, into a single asymmetry measure.
We apply the same procedures to our observed and simulated line profiles, whether or not the latter were modelled with instrumental noise. 

\subsubsection{Measuring the velocity line widths and profile edges}

To quantify the asymmetries of line profiles, it is useful to first determine their edges and widths. 
Conventional line widths, $W_{20}$ and $W_{50}$, measure the breadth of the line at 20 and 50 per cent of its peak
flux, respectively; these implicitly define the profile's low and high velocity edges, $v_{\rm l}$ and $v_{\rm h}$, respectively.
We follow a different approach and integrate the line profile to the left and to the right of its global $\HI$ flux-weighted
LOS velocity until the integrated flux reaches 95 per cent of the {\em total} flux on that side; this defines the profile edges,
 $v_{\rm l}$ and $v_{\rm h}$.
We refer to the corresponding line width as $W_{95}\equiv v_{\rm h}-v_{\rm l}$ and use this region to determine the profile's asymmetry. 
We choose $W_{95}$ because it typically traces the maximum circular velocity of a halo better than $W_{20}$ or $W_{50}$,
but has little impact on the asymmetry measures we discuss below.
The mid-point of the velocity edges are used to define the galaxy's systemic velocity,
$v_\mathrm{sys}$, i.e. $v_{\rm sys}=v_{\rm l}+0.5\,W_{95}$.

Note that for all \eagle galaxies we determine $v_{\rm l}$and $v_{\rm h}$ using their noiseless $\HI$ line profiles; we do this regardless of whether or not instrumental noise is included in estimates of their asymmetry, as the identification of edges would otherwise
be sensitive to noise. For xGASS galaxies, instrumental noise cannot be removed and, as a result, the profile edges are occasionally difficult to determine. We therefore follow \citet{Watts2020} and use the best-fit busy function \citep{Busy} as a surrogate for the line-profile shape when determining $v_{\rm l}$ and $v_{\rm h}$ (the best-fit parameters were provided by A. Watts; the fits were visually inspected for each xGASS galaxy to ensure that the best-fit busy function accurately captures the profile's edges). Their asymmetries, however, are always determined using the observed emission-line spectrum. It is worth mentioning that uncertainties on the estimated value of $v_{\rm sys}$ can lead to errors on the inferred asymmetry of line profiles, albeit small ones. For example, systematically under- or overestimating $v_{\rm sys}$ by $\pm 2$ velocity channels (corresponding to $\Delta v_{\rm sys}= 2.8\,{\rm km/s}$) typically results in $\lesssim 9$ per cent errors on each of the asymmetry measures described below.

Line profiles affected by instrumental noise -- whether simulated or observed -- may yield unreliable asymmetry measurements
if their signal-to-noise ratio ($S/N$) is not sufficiently high \citep{Watts2020,Watts2020b}. We quantify
the $S/N$ using \citep[see][]{Saintonge2007}
\begin{equation}
  S/N = \frac{1}{\sigma_\mathrm{rms}W_{95}}\sqrt{\frac{W_{95}}{2\Delta v_\mathrm{sm}}}\sum_{v=v_{\rm l}}^{v_{\rm h}}F(v)\Delta v.
\label{sbyn} 
\end{equation}
Following \citet{Watts2020}, we impose a minimum threshold of $S/N > 7$ on our simulated and observed galaxy samples. 

\subsubsection{The lopsidedness asymmetry}\label{lop}

The lopsidedness of a line profile is defined as the difference between the integrated flux on the low and high velocity
sides of $v_\mathrm{sys}$, normalized by the total flux between $v_{\rm l}$ and $v_{\rm h}$. It is defined by \citet{Peterson1974} as
\begin{equation}
  A_\textrm{l} \equiv \frac{|F_{\rm l} - F_{\rm h}|}{F_{\rm l} + F_{\rm h}},
  \label{eq:Al}
\end{equation}
where 
\begin{equation}
\displaystyle F_{\rm l} \equiv \sum_{v=v_{\rm l}}^{v_{\rm sys}}\,F(v)\Delta v,
\end{equation}
and
\begin{equation}
F_{\rm h} \equiv \sum_{v=v_{\rm sys}}^{v_{\rm h}}\,F(v)\Delta v
\end{equation}
are the total integrated fluxes between $v_{\rm sys}$ and the low- and high-velocity edges of the profile, respectively. 
This quantity is a reformulation of the ``flux ratio'', $A_{\rm fr}$, that is often used when analyzing observed
emission-line profiles \citep[e.g.][]{Haynes1998,Espada2011,Scott2018,Watts2020}.
However, unlike $A_\mathrm{fr}$, the lopsidedness is by construction confined to the range $0 \leq A_\mathrm{l} \leq 1$, with
$A_{\rm l}=0$ corresponding to a perfectly symmetric line profile.

Because $A_{\rm l}$ measures the difference in the integrated flux on the low- and high-velocity sides of
$v_{\rm sys}$, it is sensitive to global asymmetries in the $\HI$ distribution in galaxies. As a result, the line profiles of the
galaxies plotted in Fig.~\ref{illust} have very different lopsidednesses: for example, Galaxy A has $A_\mathrm{l}\approx 0$
whereas Galaxy B has a significantly higher value, $A_\mathrm{l}=0.12$. We find that \eagle galaxies for which
$A_{\rm l}\gtrsim 0.1$ (when calculated for emission-line profiles based on random lines of sight) typically have visually
disturbed $\HI$ disks.

\subsubsection{The velocity offset asymmetry}

The lopsidedness asymmetry [equation~(\ref{eq:Al})] can be calculated with respect to reference velocities other than $v_{\rm sys}$,
and there will be a particular reference velocity where $A_\mathrm{l}$ is minimized;
we refer to this as the {\em velocity of equality}, and denote it $v_\textrm{eq}$, as in \citet{Deg2020}. For a symmetric line profile,
$v_{\rm eq}=v_{\rm sys}$. Otherwise the line must be asymmetric. We therefore use the normalized offset between these two velocities as a second asymmetry
measure. Specifically, we define the velocity offset asymmetry as
\begin{equation}
  A_\mathrm{vo} = \frac{2\,|v_\textrm{sys} - v_\textrm{eq}|}{W_{95}},    
  \label{eq:Avo}
\end{equation}
which, as is the case with $A_{\rm l}$, spans the range $0\leq A_\mathrm{vo}\leq 1$.

Note that previous studies have used a similar statistic to quantify line profile asymmetries. For example,
\citet[][see also \citealt{Reynolds2020}]{Haynes1998} replaced $v_\textrm{eq}$ in equation~(\ref{eq:Avo}) with the profile's
$\HI$ flux-weighted mean velocity, $\langle v\rangle$. Although these two definitions of the velocity offset asymmetry
are quite similar, they differ in detail. As for $A_\mathrm{l}$, $A_\mathrm{vo}$ is 
most sensitive to global asymmetries. The galaxies plotted in Fig.~\ref{illust} have velocity
offset asymmetries of $A_{\rm vo}\approx 0$ (Galaxy A) and $0.16$ (Galaxy B). 

\subsubsection{The normalized residual asymmetry}

A symmetric, noiseless line profile will have the same flux in the $i^{\rm th}$ velocity bin on the lower- and higher-velocity sides of $v_{\rm sys}$, i.e., $F(v_{\rm sys}-i\Delta v)=F(v_{\rm sys}+i\Delta v$). 
Global and local asymmetries in $\HI$ line profiles should therefore manifest as
residuals in a channel-by-channel comparison of $F(v_{\rm sys}-i\Delta v)$ and $F(v_{\rm sys}+i\Delta v)$. 
We refer to this bin-wise asymmetry as the normalized
residual, which is defined as
\begin{align}
  A_\mathrm{nr} = \frac{\sum_i |F(v_{\rm sys}-i\Delta v) - F(v_{\rm sys}+i\Delta v)|\Delta v}{\sum_i [F(v_{\rm sys}-i\Delta v) + F(v_{\rm sys}+i\Delta v)]\Delta v},
  \label{eq:nres}
\end{align}
where $i$ runs from 1 to ${\rm N_{chan}/2}$ (${\rm N_{chan}}$ is the number of velocity channels between $v_{\rm l}$ and $v_{\rm h}$). As with our other asymmetry statistics, $A_{\rm nr}$ spans the range $0\leq A_{\rm nr}\leq 1$.
Unlike $A_{\rm l}$, this statistic is sensitive to small-scale asymmetries. It is analogous to the
``flipped spectrum residual'', $A_{\rm spec}$, used by \citet{Reynolds2020} to quantify the asymmetries of galaxies in the Local Volume
$\HI$ Survey \citep[LVHIS;][]{Koribalski2018}; specifically, $A_{\rm spec}=2\, A_{\rm nr}$. It is worth noting that
$A_{\rm nr}$ is sensitive to instrumental noise, and when applied to the mock line profiles of \eagle galaxies provides a
means to assess how channel-by-channel asymmetries of galaxies in observational surveys are affected by instrumental noise. Indeed, we find that our cut of ${\rm S/N}>7$ is not sufficient to fully eliminate the contribution of noise to $A_{\rm nr}$. When comparing the asymmetries of observed and simulated galaxies (Section~\ref{reconcile}) we therefore pay close attention to their distributions of ${\rm S/N}$ ratio, ensuring they overlap.

For a given galaxy, $A_{\rm nr}$ is larger than both $A_{\rm vo}$ and $A_{\rm l}$ (see Appendix~\ref{asymcor}). For example, the noiseless
$\HI$ line profile for Galaxy A (solid black lines in the lower-left panel of Fig.~\ref{illust}) has $A_\mathrm{nr}=0.03$, whereas
Galaxy B (solid black line in the lower-right panel of the same figure) has $A_\mathrm{nr}=0.35$. After adding instrumental noise
to these profiles assuming $\sigma_{\rm rms}=0.4$~mJy (i.e., the median noise level for xGASS observations), $A_{\rm nr}$
increases considerably for Galaxy A ($A_{\rm nr}=0.09$, a factor of 3 increase) but not for Galaxy B ($A_{\rm nr}=0.37$).

\subsubsection{$A_{\rm comb}$: A combined asymmetry statistic}\label{sec:Acomb}

All three asymmetry statistics described above are strongly correlated: a galaxy that appears asymmetric in any one statistic is
typically asymmetric in all three, although there are exceptions (see Appendix~\ref{asymcor}). This is because each statistic 
carries different information about the line asymmetry. It is therefore convenient to define a 
single asymmetry measure that carries their combined
information. We will refer to this as the ``combined asymmetry''; it is defined as the arithmetic mean of $A_\mathrm{l}$, $A_\mathrm{vo}$, and $A_\mathrm{nr}$, i.e. 
\begin{equation}
  A_\mathrm{comb} = \frac{A_\mathrm{l}+A_\mathrm{vo}+A_\mathrm{nr}}{3}.
  \label{eq:Acomb}
\end{equation}
Defining $A_\mathrm{comb}$ this way is preferable to using a harmonic or geometric mean since both of these result in $A_{\rm comb}\approx 0$ if any individual asymmetry measure is $\approx 0$. Conversely, the mean square weights more heavily the highest of the three asymmetry measures, which essentially is always $A_{\rm nr}$.   

We will use $A_{\rm comb}$ for the majority of our analysis but have verified that the results present in Section~\ref{results}
are qualitatively consistent for all three individual asymmetry measures. 

\section{Results}\label{results}

\subsection{Resolution requirements and sample selection}\label{resolution}

To obtain a reliable line profile for a galaxy, the position--velocity space of its $\HI$ gas must be adequately sampled; if it is not,
Poisson noise in the distribution of gas particles can result in the appearance of artificially asymmetric line profiles.
\citet{Watts2020b} refer to this effect as ``sampling-induced asymmetry'' in their analysis of 
line profiles in the TNG100 simulation \citep{Nelson2018,Pillepich2018,Springel2018}, and account
for it by imposing a lower limit on the effective number of $\HI$ resolution elements per galaxy,
$N_\mathrm{cell,\HIs}\equiv M_\mathrm{\HIs}/m_{\rm cell}$ ($M_\mathrm{\HIs}$ is the galaxy's total $\HI$ mass
and $m_{\rm cell}=1.4\times 10^6\, {\rm M}_\odot$ is the nominal cell mass for TNG100). They find that $N_\mathrm{cell,\HIs}\geq 500$
provides a reasonable compromise between galaxy statistics and accuracy of asymmetry measurements. 

To estimate the impact of particle noise on our asymmetry estimates (which may differ from the estimates of \citet{Watts2020b} due to the different hydrodynamics schemes adopted for \eagle and TNG100), we select three galaxies from the \eagle volume
and recalculate their line profiles using a randomly sampled subset of their gas particles. The number of particles
sampled varies from $\log_{10}\,N_{\rm gas}^{\rm samp}=2$ to 3.6 in equally spaced steps of
$\Delta\log_{10} N_{\rm gas}^{\rm samp}=0.1$. This is carried out a $10^3$ times for each $N_\mathrm{gas}^{\rm samp}$, and each time
we determine the corresponding number of $\HI$-rich gas particles [i.e. $N_\mathrm{gas}^{\rm samp}(f_{\HIs}>0.5)$].
In practice, we rescale the $\HI$ masses of these particles so that the galaxy's total $\HI$
remains fixed, although this does not affect the profile shape.

\begin{figure}
  \begin{center}
    \includegraphics[width=1\columnwidth]{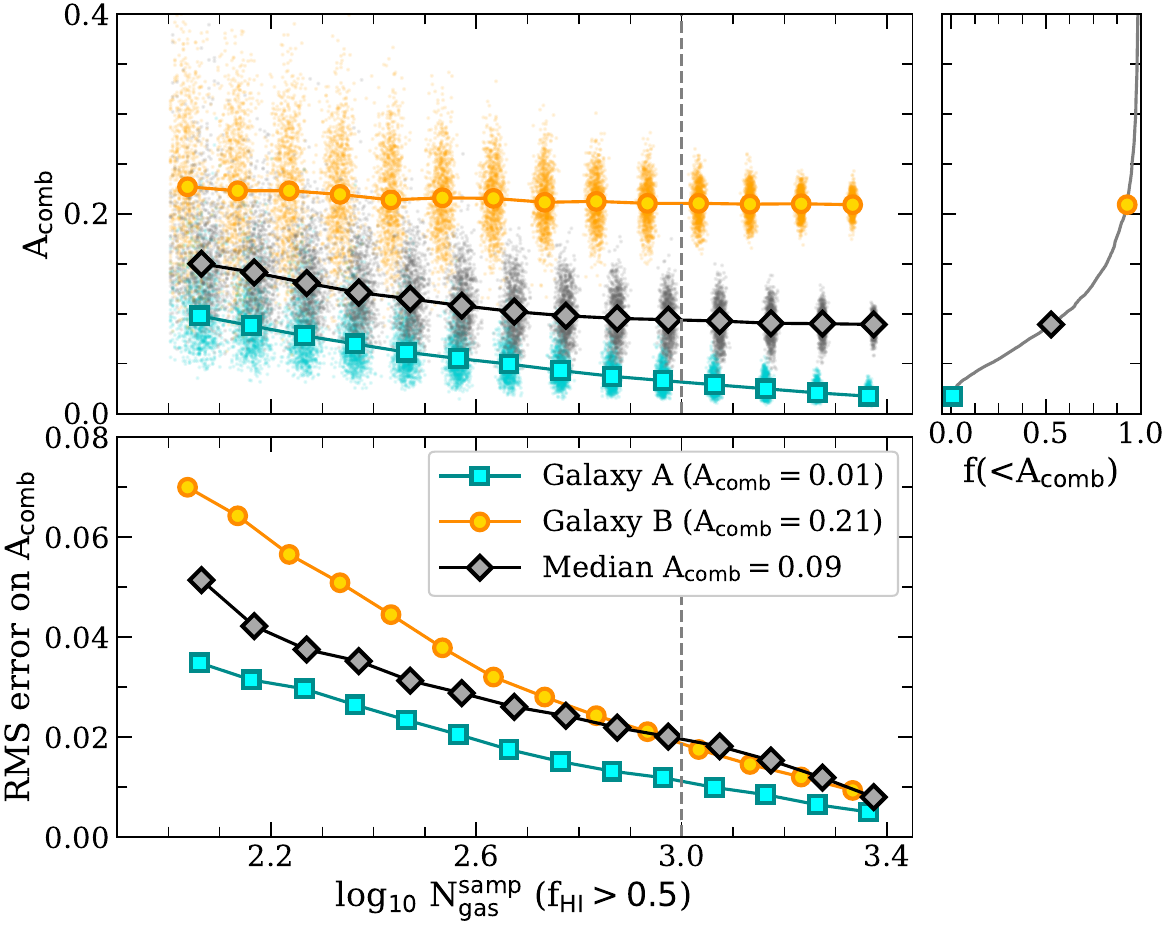}
    \caption{Impact of particle sampling on estimates of the combined asymmetry parameter, $A_{\rm comb}$ (equation~\ref{eq:Acomb}),
      for one symmetric (turquoise squares) and one asymmetric galaxy (orange circles), and for a galaxy whose line profile
      exhibits an asymmetry roughly equal to the median value of all galaxies in our \eagle sample (i.e. $A_{\rm comb}= 0.09$;
      black diamonds). For each galaxy, we randomly subsample $N^{\rm samp}_{\rm gas}$ of its gas particles (where
      $\log_{10} N^{\rm samp}_{\rm gas}$ varies from 2 to 3.6 in steps of $\Delta\log_{10}N^{\rm samp}_{\rm gas}=0.1$) and use
      them to recompute its emission line profile using the same LOS. We repeat the procedure $10^3$ times for each
      $N_{\rm gas}^{\rm samp}$ (see Section~\ref{resolution} for details). The top left panel plots the median $A_{\rm comb}$ against the median number 
      of subsampled $\HI$-rich particles, and the bottom left panel shows the resulting rms scatter among the realizations. The top
      right panel shows the cumulative histogram of $A_{\rm comb}$ for all \eagle centrals; the three coloured points correspond to
      the three galaxies used in the left-hand panels.}
    \label{pois}
  \end{center}
\end{figure}

In Fig.~\ref{pois} we plot the combined asymmetry parameter, $A_{\rm comb}$
[equation~(\ref{eq:Acomb})], obtained from the subsampled particle set as a function of the corresponding number of (subsampled) $\HI$-rich
gas particles, i.e. $N_\mathrm{gas}^{\rm samp}(f_{\HIs}>0.5)$. The coloured lines show the median asymmetries (top left panel) and
resulting rms scatter (lower panel; note that the right-most symbols in these panels correspond to the number of $\HI$-rich particles in the fully resolved
galaxies). Results are shown for one symmetric galaxy (Galaxy A from Fig.~\ref{illust}; turquoise line; $A_{\rm comb}=0.01$),
one asymmetric galaxy (Galaxy B from Fig.~\ref{illust}; orange line; $A_{\rm comb}=0.21$), and one galaxy whose combined asymmetry parameter is
approximately equal to the median value of all galaxies in our \eagle sample (i.e. $A_{\rm comb}=0.09$), which is obtained as described below.

Fig.~\ref{pois} reveals that estimates of $A_{\rm comb}$ are subject to sampling-induced asymmetry, which can lead to a subtle bias in asymmetry
estimates if galaxies are not resolved with a sufficient number of $\HI$-rich gas particles. 
However, systematic effects are small and largely independent of $N_\mathrm{gas}^{\rm samp}(f_{\HIs}>0.5)$ provided it exceeds a few hundred. Random errors on asymmetry estimates are also small provided $N_\mathrm{gas}^{\rm samp}(f_{\HIs}>0.5)$ is sufficiently large. Indeed, for all three galaxies plotted in Fig.~\ref{pois}, the rms error is below $\approx 0.02$ (lower panel) for $N_\mathrm{gas}^{\rm samp}(f_{\HIs}>0.5)\geq 10^3$, which corresponds to a typical sampling-induced
error on $A_{\rm comb}$ of $\lesssim 22$ per cent. We henceforth adopt $N_\mathrm{gas}(f_{\HIs}>0.5)\geq 10^3$ as the minimum number of $\HI$-rich
gas particles for our sample of \eagle galaxies (note that we have verified that this resolution limit also ensures our asymmetry measures are not unduly affected by the finite number of $\HI$-rich particles per velocity bin, which can vary substantially from galaxy to galaxy depending on both its $\HI$ mass and velocity line width). This threshold results in a final sample of 2924 central galaxies and 500 satellites, and the results
that follow are based on these galaxies.  

\begin{figure}
  \begin{center}
    \includegraphics[width=1\columnwidth,trim={0cm 0cm 0cm 0cm},clip=true]{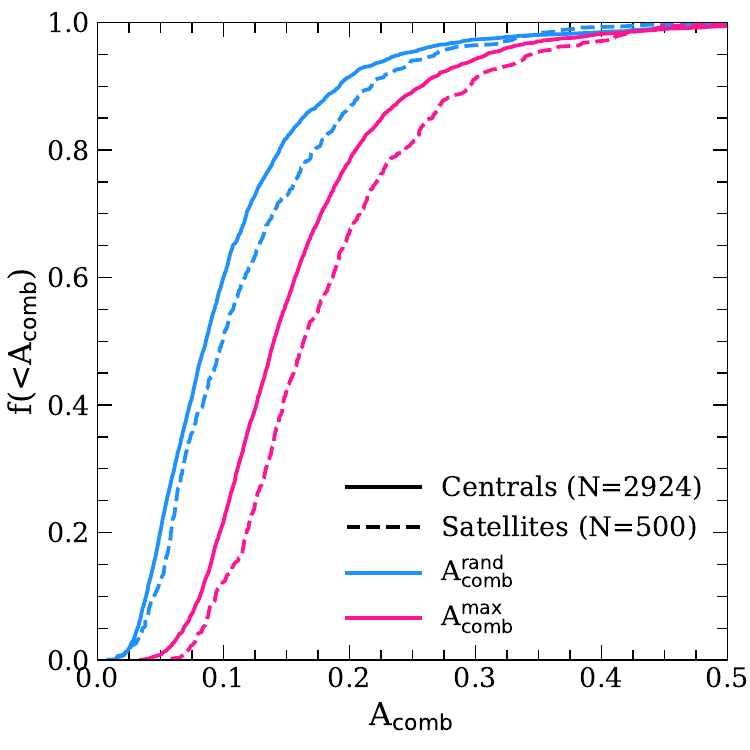}
    \caption{Cumulative distributions of $A_{\rm comb}^{\rm max}$ and $A_{\rm comb}^{\rm rand}$ for the 2924 central (solid lines)
      and 500 satellite galaxies (dashed lines)
      in the \eagle simulation that meet the selection criteria outlined in Section~\ref{resolution}.
      Blue lines correspond to line profiles constructed for
      random lines of sight; magenta lines correspond to the maximum asymmetries for the same galaxies (see
      Section~\ref{losasym} for details regarding how maximum asymmetries are calculated). In both cases,
      asymmetries are quantified using the ``combined'' asymmetry
      statistic defined in Section~\ref{sec:Acomb} (see equation~\ref{eq:Acomb}). Note that galaxies are typically more asymmetric
      than than they appear based on line profiles obtained for random viewing angles.} 
    \label{cenvsatasym}
  \end{center}
\end{figure}

\subsection{Observational effects}\label{obseff}

\subsubsection{The sensitivity of line profile shapes to line of sight}\label{losasym}

Simulations offer a unique opportunity to explore how the LOS to a galaxy impacts the shape of its $\HI$ line profile.
To investigate this, we constructed the line profiles our sample of \eagle centrals for a large number of random orientations.
The results indicate that -- provided low inclinations are avoided, for which line profiles are narrow, and asymmetry estimates
unreliable -- asymmetries tend to be highest when galaxies are viewed nearly edge-on, i.e. for particular lines of sight that are
perpendicular to the angular momentum vector of the $\HI$ disk, $\vec{L}_{\HIs}$. This is because the profile
width -- and therefore the number of velocity channels by which $W_{95}$ is resolved -- increases with inclination, and the line
profile better samples the global velocity structure of the disk. Hence, we may estimate the {\em maximum} asymmetries
of \eagle galaxies by orienting them edge-on and generating $\HI$ profiles for a range of viewing angles spanning
$0^\circ\leq\phi\leq180^\circ$, where $\phi$ is the angle by which the galaxy is rotated about $\vec{L}_{\HIs}$.
In practice, we vary $\phi$ in equally spaced steps of $\Delta\phi=5^\circ$, resulting 36 edge-on views for each galaxy.
Note that the $\phi=0$ viewing angle is arbitrary; note also that we do not consider $\phi \geq 180^\circ$ because, for edge-on orientations, the profiles for $\phi$ and $\phi+180^\circ$ are inversions of each other in velocity space and therefore have equal asymmetries.

This procedure was carried out for all central and satellite galaxies that meet our resolution criterion
in order to determine the maximum value of their combined asymmetry, $A_\mathrm{comb}^{\rm max}$;
the cumulative distributions are plotted in Fig.~\ref{cenvsatasym} as solid and dashed magenta lines, respectively. 
For comparison, the distribution of $A_\mathrm{comb}^{\rm rand}$ obtained for random 
orientations\footnote{To obtain random orientations for galaxies, we rotate their $\vec{L}_{\HIs}$'s such that the corresponding unit vectors uniformly 
sample a unit shell.} are shown as solid and
dashed blue lines for centrals and satellites, respectively. (We hereafter distinguish asymmetry measurements obtained for
random viewing angles from those that maximize the inferred emission line asymmetry using superscripts; for example,
$A_{\rm comb}^{\rm rand}$ is obtained for random viewing angles. If orientation is not relevant, we drop the superscript.)

There are a few points worth highlighting in Fig.~\ref{cenvsatasym}. One is that the {\em maximum} line profile asymmetries
obtained for edge-on orientations are considerably larger than those obtained for random orientations. For example, the
median asymmetry increases from $A_{\rm comb}^{\rm rand}\approx 0.09$ to $A_{\rm comb}^{\rm max}\approx 0.14$
(for satellite galaxies the medians are $A_{\rm comb}^{\rm rand}\approx 0.1$ and
$A_{\rm comb}^{\rm max}\approx 0.16$).
Note too that very few galaxies are {\em intrinsically} symmetric: For example, less than 22 per cent of centrals (and less than 13 per cent of
satellite galaxies) have $A_{\rm comb}^{\rm max}\leq 0.1$, whereas 60 per cent have $A_{\rm comb}^{\rm rand}\leq 0.1$ (51 per cent for
satellites). Note also, in agreement with \citet{Watts2020b}, that satellite galaxies typically exhibit higher asymmetries than centrals. As discussed in Section~\ref{satellites}, this is likely a result of the different environmental processes that affect the dynamical evolution of centrals and satellites.

\subsubsection{The impact of effective velocity resolution, instrumental noise, and galaxy distance on line profile asymmetries}\label{edgeasym}

\begin{figure}
  \begin{center}
    \includegraphics[width=0.91\columnwidth,trim={0cm 0.1cm 0cm 0cm},clip=true]{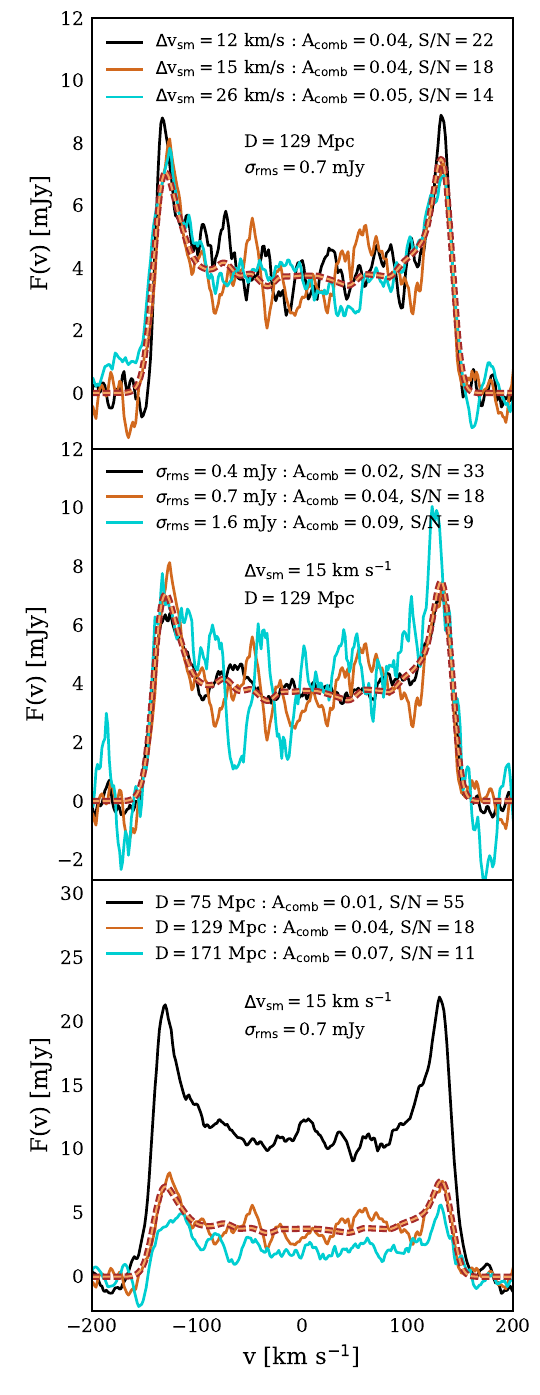}
    \caption{From top to bottom, panels show the impact of effective velocity resolution ($\Delta v_{\rm sm}$), instrumental noise ($\sigma_{\rm rms}$),
      and galaxy distance ($D$) on the shape of the $\HIs$ emission-line profile of a simulated galaxy whose noiseless
      profile has a combined asymmetry $A_{\rm comb}=0.01$. We choose fiducial values for $\Delta v_{\rm sm}$
      ($15\,{\rm km\,s^{-1}}$), $\sigma_{\rm rms}$
      ($0.7\,{\rm mJy}$) and $D$ ($129\,{\rm Mpc}$) that coincide with the median values for galaxies in the xGASS
      survey; in each panel, two of these are held fixed while the third is varied such that it spans the interquartile
      range of xGASS observations (by varying the parameters this way, the brown line is repeated in each panel). For each
      combination of parameters, we quote the combined asymmetry of the resulting line, $A_{\rm comb}$ (equation~\ref{eq:Acomb}),
      and its signal-to-noise ratio, $S/N$ (equation~\ref{sbyn}). In each panel, the noiseless profile for $\Delta v_{\rm sm}=15\,{\rm km\,s^{-1}}$ and $D=129\,{\rm Mpc}$ is shown as the dashed curve for comparison.}
    \label{obillust}
  \end{center}
\end{figure}

For a particular viewing angle there are additional factors that can contribute to the observed
asymmetry of a galaxy's emission-line profile: the effective velocity resolution ($\Delta v_{\rm sm}$) and instrumental noise ($\sigma_{\rm rms}$)
associated with the line profile, and the
galaxy's distance [$D$; see equation~(\ref{profile})]. In Fig.~\ref{obillust} we illustrate the impact of each of these on an underlying symmetric profile (i.e. Galaxy A, Fig.~\ref{illust}). 
To do so, and to make connection with xGASS
observations, we chose fiducial values of the above parameters that coincide with the median values for galaxies in the xGASS survey:
$\Delta v_{\rm sm}=15\,{\rm km\,s^{-1}}$, $\sigma_{\rm rms}=0.7\,{\rm mJy}$, and $D=129\,{\rm Mpc}$. The resulting line
profiles are plotted as brown curves in each panel of Fig.~\ref{obillust}, for reference.
The remaining curves in each panel show the effect of varying one of these parameters while the other two are held
fixed: in the upper panel we vary $\Delta v_{\rm sm}$ from $12\,{\rm km\,s^{-1}}$ to $26\,{\rm km\,s^{-1}}$; in the
middle panel we vary $\sigma_{\rm rms}$ from $0.4\,{\rm mJy}$ to $1.6\,{\rm mJy}$; and in the bottom panel $D$
from $75$ to $171\,{\rm Mpc}$. Note that these parameter ranges are not chosen arbitrarily, but to span the interquartile
range of xGASS observations.

In general, increasing any of these parameters leads to a lower $S/N$, as indicated in the legend of Fig.~\ref{obillust}. Although increasing the
effective velocity resolution $\Delta v_{\rm sm}$ from 12 ${\rm km\,s^{-1}}$ to 26 ${\rm km\,s^{-1}}$ reduces $S/N$ by a factor of $\approx 1.6$, the combined asymmetry parameter increases only a little, from $A_{\rm comb}=0.04$ to $0.05$. This is primarily due to $A_{\rm l}$, which increases from 0.03 to 0.06, while $A_{\rm vo}$ and $A_{\rm nr}$ are less affected, changing from 0.02 to 0.03 and 0.08 to 0.07, respectively.
The middle panel shows that the line asymmetry is more sensitive to instrumental noise: $A_\mathrm{comb}$ changes
from $0.02$ for $\sigma_\mathrm{rms}=0.4$~mJy to $0.09$ for $\sigma_\mathrm{rms}=1.6$~mJy (the corresponding $S/N$ ratio is reduced by a factor of
$\approx 4$). The corresponding changes in $A_{\rm l}$ and $A_{\rm nr}$ are of similar magnitudes, 
with $A_{\rm l}$ increasing from 0.01 to 0.09 and $A_{\rm nr}$ from 0.05 to 0.14. $A_{\rm vo}$ incurs a smaller change, from $\approx 0$ to $0.04$. 
A more distant galaxy (of a given $\HI$ mass) has a lower observed $\HI$ flux which reduces the $S/N$ for
a given $\sigma_{\rm rms}$; this typically results in a higher asymmetry. For example, in the bottom panel of Fig.~\ref{obillust}, the $S/N$ is reduced by factor of $\approx 5$, and $A_\mathrm{comb}$ increases from $0.01$ to $0.07$ when the galaxy's assumed distance increases from $D=75$~Mpc to $D=171$~Mpc. $A_{\rm l}$ 
and $A_{\rm vo}$ increase from 0.01 to 0.04 and 0.03, respectively, whereas $A_{\rm nr}$ is impacted the most with an increase from 0.03 to 0.15. Similar results also apply to other galaxies in our sample, suggesting that it is important to carefully consider observational effects when comparing simulated line profiles to observed ones.

\subsection{A comparison of line profile asymmetries in \eagle and xGASS}\label{reconcile}

\begin{figure*}
  \begin{center}
    \includegraphics[width=1.4\columnwidth,trim={0cm 0cm 0cm 0cm},clip=true]{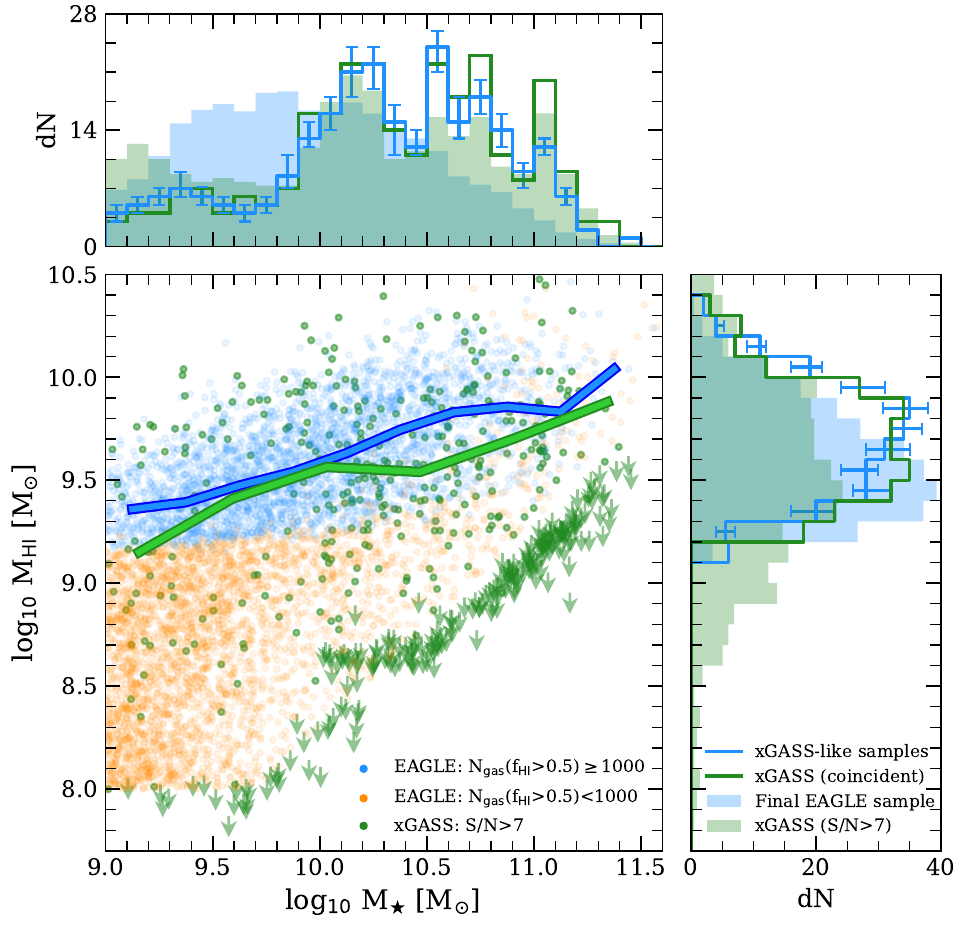}
    \caption{Total $\HI$ mass versus stellar mass for central galaxies in the xGASS survey and the \eagle simulation. The blue points show the \eagle galaxies with $M_\star\geq 10^9\,{\rm M_\odot}$ that also contain at least $10^3$ gas particles with $\HI$ fractions $f_{\HIs}\geq 0.5$; those with $\HI$ fractions above the xGASS detection limit but with fewer than $10^3$ $\HI$-rich particles are shown in orange. Green points show the xGASS galaxies whose $\HI$ line profiles have signal-to-noise ratios $S/N\geq 7$ (and are resolved by $>20$ velocity channels); the downward pointing arrows are xGASS non-detections. The blue and green curves in the main panel plot the medians of the \eagle sample [$N_{\rm gas}(f_{\HIs} \geq 0.5)\geq 10^3$; i.e. the blue points] and the xGASS sample ($S/N\geq 7$; i.e. the green points), respectively. Panels on the top and right-hand side show the distributions of the $\HI$ masses and stellar masses, respectively, for various samples of galaxies used in our analysis. The filled blue histograms correspond to our final sample of \eagle galaxies with $M_\star\geq 10^9\,{\rm M_\odot}$ and $N_\mathrm{gas}(f_{\HIs}>0.5) \gtrsim 10^3$ (see Section~\ref{resolution} for details); filled green histograms correspond to all xGASS galaxies whose line profiles have $S/N>7$. The open green histograms correspond to xGASS galaxies with $S/N>7$ that overlap with our final \eagle sample in the $M_{\HIs}$-$M_\star$ plane, and the open blue histograms correspond to xGASS-like subsamples drawn from \eagle (see Section~\ref{reconcile} for details). The error bars on the open blue histograms in the upper and right-hand panels show $20^{\rm th}$--$80^{\rm th}$ percentile scatter for the subsamples. Note that all histograms have been normalized to have the same area.}
    \label{mhivsmstar}
  \end{center}
\end{figure*}

Although projection effects, effective velocity resolution, instrumental noise and distance all affect the inferred asymmetry of a galaxy's $\HI$ emission-line profile, it is nevertheless possible to meaningfully compare the line profiles of simulated and observed systems.
However, care must be taken to ensure that differences in their selection criteria -- which may ultimately result in a very different distribution of galaxies in the stellar mass-$\HI$ mass plane -- do not unduly bias their asymmetry distributions.

In Fig.~\ref{mhivsmstar} we plot the relation between total $\HI$ mass and stellar mass for galaxies in
\eagle and xGASS. Only xGASS galaxies whose line profiles were observed with $S/N\geq 7$ (and resolved by $>20$ velocity channels)
are included in the analysis that follows; these are shown as green circles in Fig.~\ref{mhivsmstar}
(downward-pointing green arrows indicate non-detections). \eagle galaxies are shown as blue points if
$N_\mathrm{gas}(f_{\HIs}>0.5) \geq 10^3$ and as orange points otherwise. The blue and green filled histograms
in the upper and right-hand panels of Fig.~\ref{mhivsmstar} indicate the distributions of stellar and
$\HI$ masses for \eagle galaxies that meet our resolution cut and for xGASS galaxies, respectively
(note that the blue histogram has been normalized to have the same area as the green one). 

\begin{figure}
  \begin{center}
    \includegraphics[width=1\columnwidth,trim={0cm 0cm 0cm 0cm},clip=true]{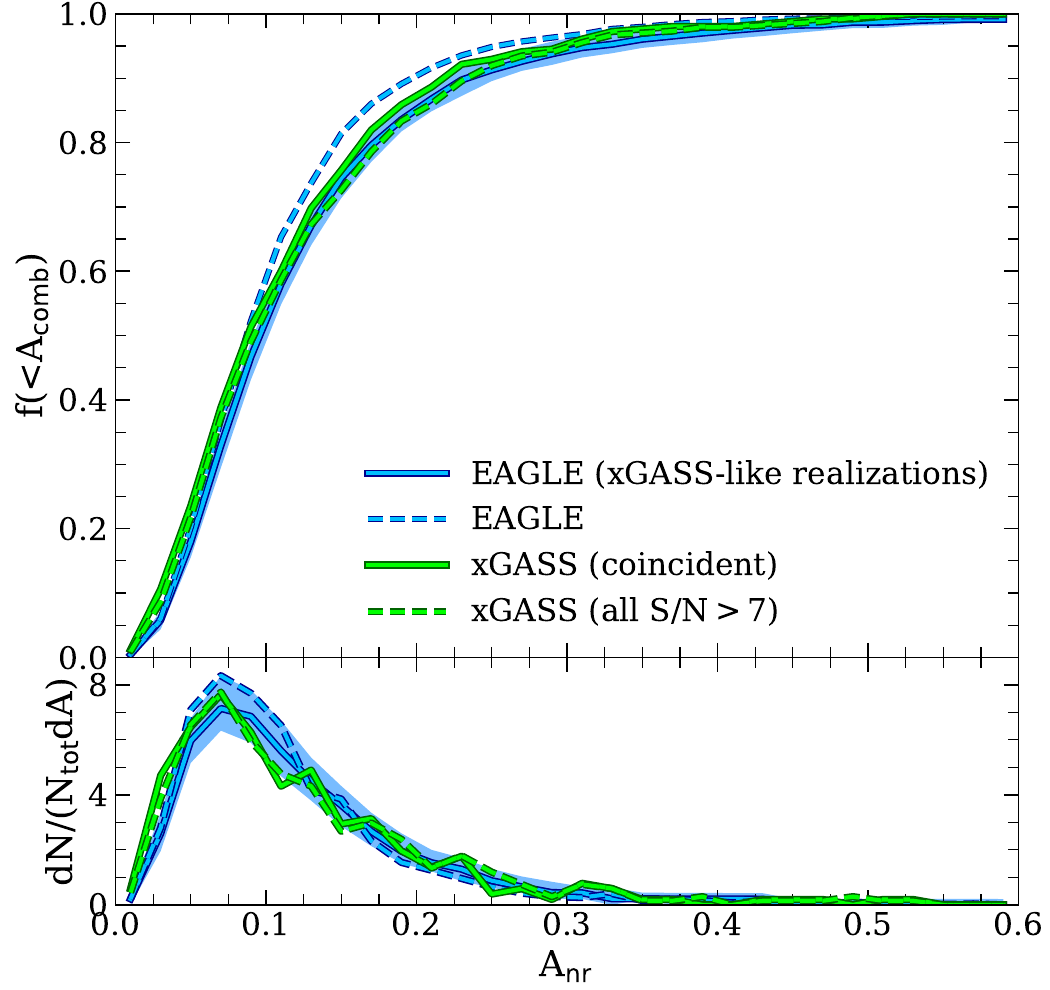}
    \caption{Distribution of line profile asymmetries for central galaxies in xGASS ($S/N\geq 7$) and for xGASS-like subsamples of central galaxies drawn from the \eagle simulation (see Section~\ref{reconcile} and Fig.~\ref{mhivsmstar} for details). The top panel plots the cumulative distribution of $A_{\rm comb}^{\rm rand}$ (see equation~\ref{eq:Acomb}); the bottom panel shows the differential distribution. The dashed blue curve corresponds to the distribution of $A_{\rm comb}^{\rm rand}$ for {\em all} \eagle galaxies with $M_\star\geq 10^9\,{\rm M_\odot}$ that also pass our resolution criterion, i.e. $N_\mathrm{gas}(f_{\HIs}>0.5) \geq 10^3$; the dashed green
      curve corresponds to all central galaxies in xGASS. Note that, although these distributions are
      similar, the slight offset is primarily driven by differences in how \eagle and xGASS galaxies
      populate the $M_{\HIs}-M_\star$ space (see Fig.~\ref{mhivsmstar}). The solid blue curve, for example, corresponds to median
      asymmetry distribution for 100 xGASS-like realizations of \eagleNS; the shaded region indicates the corresponding $20^\mathrm{th}$--$80^\mathrm{th}$ interpercentile 
      range for those realizations. 
      Note that this agrees much better with the solid green curve, which shows
      the asymmetry distributions of ($S/N\geq 7$) xGASS centrals that overlap with the \eagle galaxies in the $M_{\HIs}-M_{\star}$ plane.}
    \label{xgasymcomp}
  \end{center}
\end{figure}

Note that the distributions of $M_{\HIs}$ and $M_\star$ differ considerably between \eagle and xGASS, with \eagle
galaxies strongly biased toward lower stellar masses, and slightly biased to higher $\HI$ masses. When comparing asymmetries
we account for these differences -- which are primarily the result of our resolution requirement (Section~\ref{resolution}) and the xGASS
survey selection -- by sampling populations of galaxies from \eagle that mimic the distributions of the
$\HI$ and stellar masses in xGASS. This is achieved by dividing the $M_{\HIs}-M_\star$ plane into 
a grid comprised of 16 horizontal bins logarithmically spaced between $\log_{10}(M_\star/\rm{M}_\odot)=[9,11.5]$,
and 18 vertical bins logarithmically spaced between $\log_{10}(M_{\HIs}/\rm{M}_\odot)=[7.7,10.5]$. In each grid cell, we
randomly sample the same number of \eagle galaxies as there are galaxies in xGASS, or all the \eagle galaxies in the cell,
whichever is fewer. This sampling is performed 100 times to generate 100 xGASS-like realizations of \eagle galaxies, which is sufficient to
ensure that all simulated galaxies are selected at least once. If a particular \eagle galaxy appears in multiple realizations -- for
example, in grid cells more densely populated by galaxies in xGASS than in \eagle -- we use a different, random viewing
angle to construct its line profile each time. xGASS galaxies that occupy a grid cell in which there are none in \eagle
are excluded from the asymmetry analysis below (a total of 79 galaxies are removed from the xGASS sample as a result).
The open green and blue histograms in the upper and right-hand panels of Fig.~\ref{xgasymcomp} show the
resulting distributions of $\HI$ and stellar mass for resulting xGASS and \eagle samples, normalized to have the same
area as the filled histograms (note that the error bars on the blue histograms reflect the variation among the 100
xGASS realizations of \eagle galaxies).

As discussed in Section~\ref{obseff}, the distance to a galaxy and the instrumental noise associated with its line profile will ultimately affect its inferred asymmetry. When comparing the $\HI$ line asymmetries of xGASS galaxies to those of simulated \eagle galaxies it is therefore important to choose sensible distances and noise levels for the latter. xGASS consists of two distinct galaxy populations: one corresponding to stellar masses $M_\star\lesssim 10^{10}\,{\rm M}_\odot$ for which the median distance, noise and effective 
resolution are $D=71$~Mpc, $\sigma_{\rm rms}=1.3$~mJy and $\Delta v_{\rm sm}=12~\kms$, respectively; and another for $M_\star\gtrsim 10^{10}\,{\rm M}_\odot$, for which the corresponding medians are $D=162$~Mpc, $\sigma_{\rm rms}=0.5$~mJy and $\Delta v_{\rm sm}=15~\kms$. When assigning these parameters to \eagle galaxies we likewise separate them into stellar mass bins above and below $10^{10}\,M_\odot$, and assign to them the median $D$, $\sigma_{\rm rms}$ and $\Delta v_{\rm sm}$ of the corresponding xGASS population. These profiles, and those of xGASS centrals, are then analyzed as described in Section~\ref{measure} in order to determine their
asymmetries. The solid blue and green lines in the upper panel of Fig.~\ref{xgasymcomp} compare the cumulative
probability distributions of $A_{\rm comb}^{\rm rand}$ obtained for \eagle and xGASS, respectively; the corresponding lines
in the lower panel compare the differential distributions. The interquartile scatter among all
100 realizations of \eagle galaxies is indicated using a shaded blue region. Note the excellent agreement between the asymmetry distributions of \eagle and xGASS galaxies when constructed this way. For comparison, the dashed blue and green
lines in each panel show the corresponding distributions of $A_{\rm comb}^{\rm rand}$
for {\em all} \eagle galaxies [with $N_\mathrm{gas}(f_{\HIs}>0.5) \geq 10^3$] and {\em all} xGASS galaxies (with $S/N\geq 7$),
without attempting to match their distributions in $\HI$ or stellar mass. Although these asymmetry distributions are similar,
it is clear that appropriately sampling the distributions of $\HI$ and stellar mass for \eagle galaxies (as well as assigning sensible distances and instrumental noise) improves the overall agreement.\footnote{A Kolmogorov-Smirnov test on the asymmetry distributions of all \eagle galaxies and all (${\rm S/N}>7$) xGASS galaxies suggests that the probability of them being drawn from the same underlying distribution function is only $\approx 10$ per cent. When carefully matching the distributions of \eagle and xGASS galaxies in stellar and $\HI$ mass (as described in Section~\ref{reconcile}) the probability increases to $\approx 72$ per cent.}

\subsection{The relationship between $\HI$ emission-line asymmetry and the properties of central galaxies and their host dark matter haloes}\label{centrals}

Having established that the asymmetries of the $\HI$ emission lines of \eagle galaxies sensibly reproduce the observed asymmetries of galaxies in xGASS,
we next turn our attention to the physical drivers of asymmetry. To do so, we focus our analysis on the noiseless $\HI$ emission-line profiles of \eagle galaxies, so that our asymmetry estimates are not subject to the potential observational uncertainties
described in Section~\ref{edgeasym}. For each galaxy, we consider the maximum value of its $\HI$ asymmetry parameter, i.e.
$A_{\rm comb}^{\rm max}$, as well as $A_{\rm comb}^{\rm rand}$ determined for one random viewing angle.

Fig.~\ref{AsymMstar} plots $A_{\rm comb}^{\rm max}$ (upper panel) and $A_{\rm comb}^{\rm rand}$ 
(lower panel) versus galaxy stellar mass, $M_\star$. Grey points in each panel show the full population of \eagle
galaxies that pass our resolution cut; the medians and $20^\mathrm{th}-80^\mathrm{th}$ percentile scatter are shown using
outsized circles and error bars, respectively. Both asymmetry estimates are largely independent of $M_\star$, except perhaps at the
highest masses, where both $A_{\rm comb}^{\rm max}$ and	$A_{\rm comb}^{\rm rand}$ increase slightly. The trends, however, are
quite weak: galaxies with $M_\star\gtrsim 10^{10.5}\,\rm{M}_\odot$, for example, have average asymmetries that are only $\approx 19$ per cent larger than those of the full
galaxy population. A similar relation between asymmetry and stellar mass was also reported by \citet{Watts2020}, who compared the asymmetries
of xGASS galaxies above and below $M_\star=10^{10}$~M$_\odot$ and found no systematic difference.

In the following we will exploit the approximate mass-independence of galaxy asymmetries
to explore how other galaxy properties -- such as their $\HI$ and stellar morphologies, gas fractions, star formation rates, accretion rates, and
the dynamical state of their dark matter haloes -- differ between the populations of galaxies with highly symmetric,
or highly asymmetric $\HI$ reservoirs. To do so, we compare and contrast the properties of galaxies and halos that
harbour the most symmetric or asymmetric systems. Specifically, we identify the galaxies that, at fixed $M_\star$, occupy the regions below 20$^\mathrm{th}$ and above 80$^\mathrm{th}$ percentile of the asymmetry distribution. These are shown as brown and blue points 
in each panel of Fig.~\ref{AsymMstar}, respectively. We will use these galaxy subsamples in the next few sections of the paper.

\subsubsection{The $\HI$ fractions, star formation rates, and gas flows in galaxies with symmetric and asymmetric $\HI$ emission lines}\label{gasflows}

\begin{figure}
  \begin{center}
    \includegraphics[width=1\columnwidth,trim={0cm 0cm 0cm 0cm},clip=true]{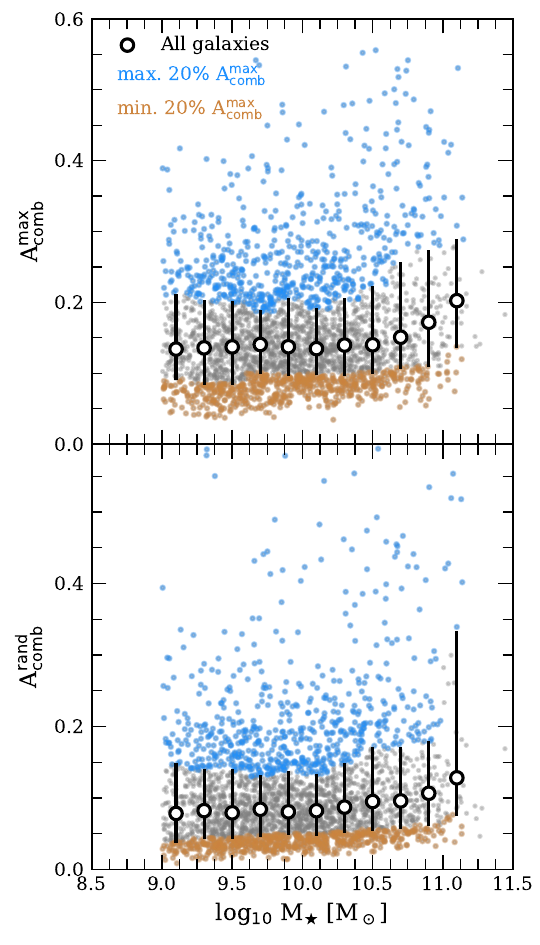}
    \caption{The combined asymmetry parameters [$A_{\rm comb}$; equation~(\ref{eq:Acomb})] versus galaxy stellar mass for
      all \eagle centrals resolved with at least $10^3$ $\HI$-rich gas particles
      [i.e. $N_\mathrm{gas}(f_{\HIs}>0.5) \geq 10^3$]. The upper and lower panels respectively 
      correspond to the maximum asymmetry for each galaxy ($A_{\rm comb}^{\rm max}$) and a value obtained for a
      random viewing angle ($A_{\rm comb}^{\rm rand}$).
      Our full sample of 2924 \eagle galaxies is shown using light grey points; the median and $20^\mathrm{th}-80^\mathrm{th}$ percentile
      scatter are shown using white circles and error bars, respectively. The blue and brown points in each
      panel indicate the upper and lower 20 per cent region of $A_{\rm comb}^{\rm max}$ (upper panel) and
      $A_{\rm comb}^{\rm rand}$ (lower panel) in each stellar mass bin, respectively. Both the maximum and combined asymmetry of \eagle
      galaxies are largely independent of $M_\star$ except, perhaps, for the most massive galaxies, for
      which there are hints of slightly higher asymmetries. }
    \label{AsymMstar}
  \end{center}
\end{figure}

Late-type galaxies, as well as galaxies in close pairs, are known to exhibit a higher incidence of global asymmetry
than populations as a whole \citep{Haynes1998,Matthews1998,Espada2011}. Late types are gas rich \citep{Wang2013}, and post-merger galaxies of a given
stellar mass have higher than average $\HI$ fractions \citep{Ellison2018}. Although circumstantial, this is evidence that
emission-line asymmetries may be related to stochastic gas accretion or mergers, since asymmetric line profiles
appear more common in gas-rich galaxies. These ideas, however, were recently challenged by \citet{Reynolds2020b}, who found
no relation between gas content and the asymmetry of 21-cm emission lines for galaxies in the HIPASS survey \citep{HIPASS}.
\citet{Watts2021} showed that, for galaxies in xGASS, asymmetries are in fact more common in {\em gas-poor} galaxies, and how survey selection effects can give rise to the apparent conflict between prior observational studies. 

\begin{figure}
  \begin{center}
    \includegraphics[width=0.97\columnwidth,trim={0cm 0cm 0cm 0cm},clip=true]{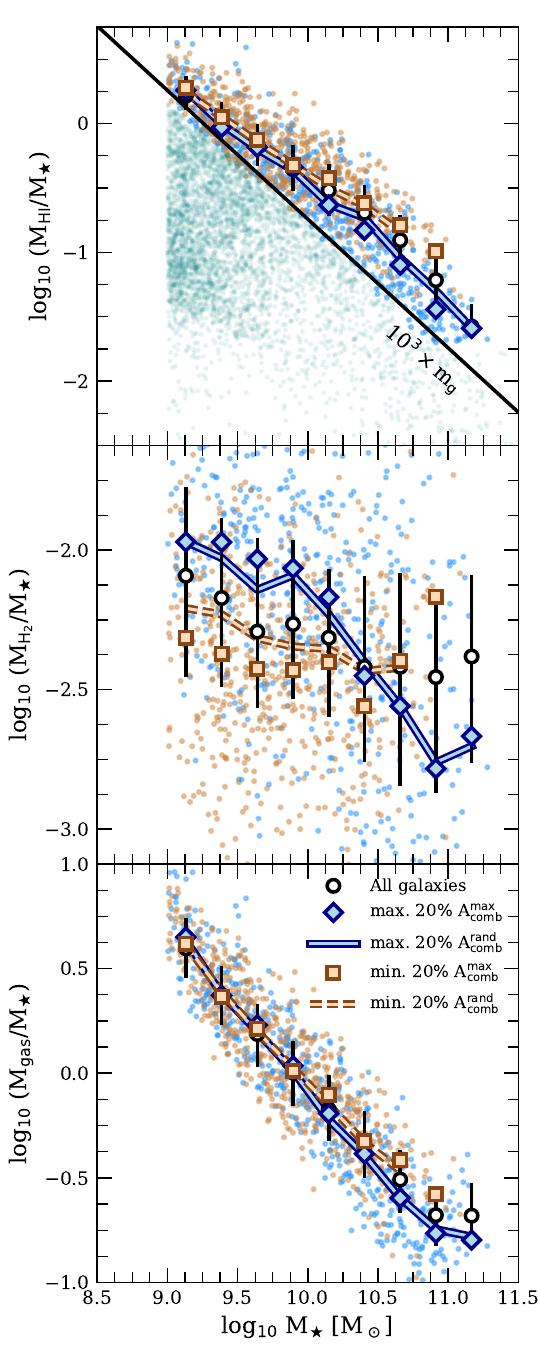}
    \caption{The $\HI$ (upper panel), ${\rm H_2}$ (middle panel) and total gas content (lower panel) of \eagle galaxies plotted as a function of their stellar mass (to reduce the dynamic range we have normalized all quantities by the galaxy's stellar mass). The white circles and error bars in each panel indicate the median and $20^{\rm th}$--$80^{\rm th}$ interpercentile range of the entire sample of \eagle galaxies that possess at least $10^3$ $\HI$-rich gas particles. As in the upper panel of Fig.~\ref{AsymMstar}, the blue and brown points in each panel correspond to galaxies that, at a given stellar mass, occupy the upper and lower 20 per cent of maximal asymmetries, $A_{\rm comb}^{\rm max}$; their median relations are shown using blue diamonds and brown squares, respectively. For comparison, the median trends based on the upper and lower 20 per cent of $A_{\rm comb}^{\rm rand}$ are shown using solid blue and dashed brown lines, respectively. The teal points in the upper panel are galaxies that do not pass our resolution cut of $\geq 10^3$ $\HI$-rich gas particles.}
    \label{Asymfgas}
  \end{center}
\end{figure}

We explore the link between the $\HI$ content of \eagle galaxies and their $\HI$ line asymmetries in the
top panel of Fig.~\ref{Asymfgas}. Here we plot the $\HI$ fraction (i.e. the $\HI$-to-stellar mass ratio)
versus stellar mass for all \eagle galaxies that are resolved with at least $10^3$ $\HI$-rich gas particles (teal points correspond to galaxies that do not pass our resolution criterion -- these are not included in
the analysis that follows). White circles and errors bars
indicate the median trends and the $20^\mathrm{th}-80^\mathrm{th}$ percentile scatter, respectively. Blue and brown points correspond to the
$\HI$ fractions of galaxies in the upper and lower 20$^\mathrm{th}$ percentile of $A_{\rm comb}^{\rm max}$, respectively
(i.e., for the same sets of colored points shown in the upper panel of Fig.~\ref{AsymMstar});
their corresponding medians in bins of stellar mass are shown using blue diamonds and brown squares, respectively.
The trends suggest that \eagle galaxies with asymmetric $\HI$ line profiles have, on average, less-massive $\HI$ reservoirs than those
of symmetric systems, in agreement with the conclusions of \citet{Watts2021}. Note however that the difference in the median $\HI$ fractions of 
these two samples is largest for the most massive galaxies in our sample, but weakens substantially towards lower stellar masses. This result however is
unlikely physical: it is driven, at least in part, by the resolution cut imposed on our sample of \eagle galaxies
(shown as a solid black line in Fig.~\ref{Asymfgas}, for reference).
Note that the slight difference in the average gas fractions is also present for asymmetries based on random viewing angles which, as shown in Section~\ref{losasym},
systematically underestimate the intrinsic asymmetries of galaxies. The solid blue and dashed brown lines in
the upper panel of Fig.~\ref{Asymfgas}, for example, show the median relations for the upper and lower 20$^\mathrm{th}$
percentiles of $A_{\rm comb}^{\rm rand}$.

Interestingly, unlike their $\HI$ fractions, we find that asymmetric galaxies with stellar masses below about $10^{10.3}\,{\rm M_\odot}$ possess
considerably more molecular hydrogen than symmetric systems of similar stellar mass, a result shown explicitly in the middle panel of
Fig.~\ref{Asymfgas}. This is true despite both galaxy subsamples having comparable {\em total} gas content below $\approx 10^{10.3}\,{\rm M_\odot}$ (the lower-most panel of Fig.~\ref{Asymfgas} plots the total gas mass fraction versus stellar mass). Above $\approx 10^{10.3}\,{\rm M_\odot}$, galaxies with asymmetric $\HI$ lines have lower $\HI$ and ${\rm H}_2$ reservoirs, which is likely a result of their lower-than-average total gas content (lower-most panel of Fig.~\ref{Asymfgas}). 

\begin{figure}
  \begin{center}
    \includegraphics[width=1\columnwidth,trim={0cm 0cm 0cm 0cm},clip=true]{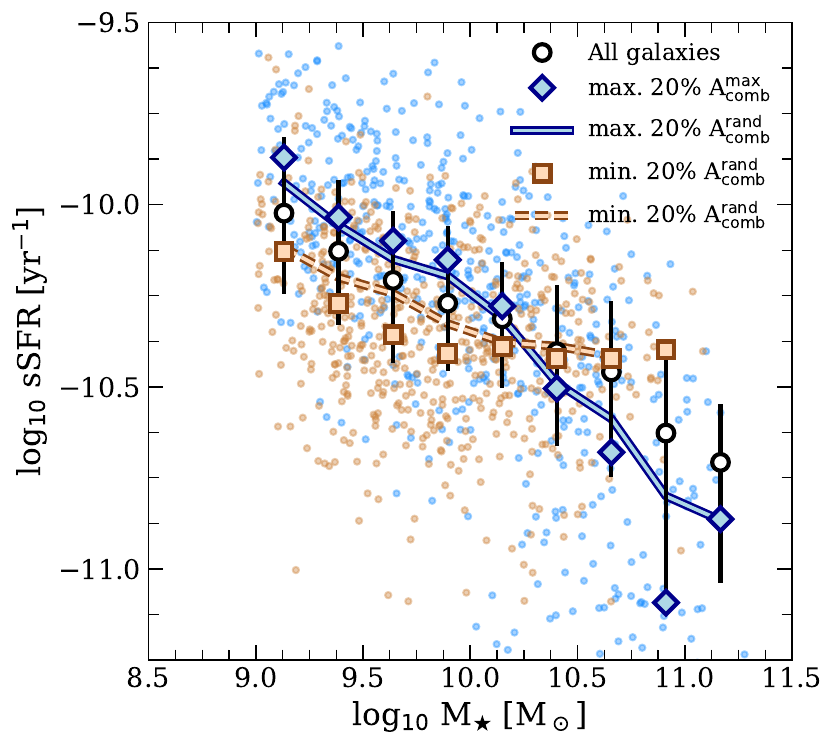}
    \caption{The relation between specific star formation rate (sSFR) and stellar mass for \eagle galaxies (resolved with $\gtrsim 10^3$ $\HI-$rich particles) with symmetric (brown) and asymmetric (blue) $\HI$ emission-line profiles. Plotting conventions follow Fig.~\ref{Asymfgas}.}
    \label{ssfrvsmstar}
  \end{center}
\end{figure}

Based on systematic differences in the observed $\HI$ content of galaxies with symmetric and asymmetric line profiles, \citet{Watts2021}
suggested that physical process capable of disturbing or removing $\HI$ gas from galaxies may be the primary drivers of
$\HI$ line asymmetry. Indeed, \citet{Watts2021} note that galaxies with the highest asymmetries also exhibit elevated star
formation rates, which may be indicative of feedback-driven gas removal in those systems. It is therefore unsurprising that specific star formation
rates (sSFR) differ between \eagle galaxies with the highest and lowest global $\HI$ line asymmetries. We show this explicitly
in Fig.~\ref{ssfrvsmstar}, where we plot our sample of \eagle galaxies in the
sSFR--stellar mass plane (the various symbols, colours and lines have the same meaning as in Fig.~\ref{Asymfgas}); the SFRs have been averaged over $0.5$~Gyr. 
Below $M_\star\approx 10^{10.3}\,{\rm M}_\odot$, galaxies with asymmetric global $\HI$ spectra exhibit
elevated star formation rates relative to those with symmetric spectra; above $M_\star\approx 10^{10.3}\,{\rm M}_\odot$,
their star formation rates are suppressed relative to those with symmetric spectra. Note that the differences in the mass-dependence of the specific star formation
rates of symmetric and asymmetric \eagle galaxies are qualitatively similar to the differences in their ${\rm H}_2$ fractions.
This is not surprising: both star formation in \eagle and the molecular hydrogen content of gas particles (calculated in post-processing; see Section~\ref{himod}), increase with increasing gas density and metallicity, so gas particles with the highest star formation
rates are also the ones predicted to have the most ${\rm H_2}$. 

\begin{figure}
  \begin{center}
    \includegraphics[width=1\columnwidth,trim={0cm 0cm 0cm 0cm},clip=true]{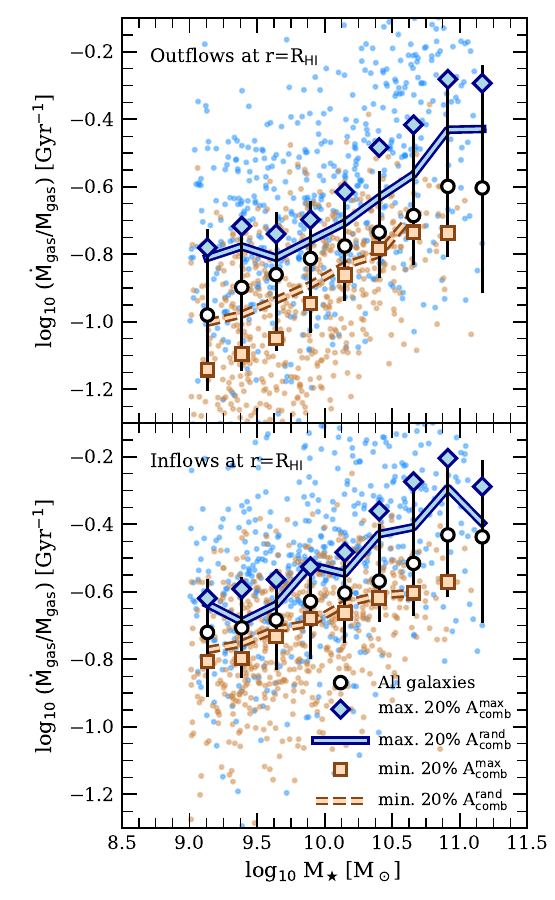}
    \caption{The outflow (upper panel) and inflow (lower panel) rates of gas particles averaged over last $\approx 0.6$~Gyr. Both rates are based on the total flux of gas particles that cross a spherical aperture of physical radius $r=R_{\HIs}$ surrounding the main progenitor of each central galaxy in our \eagle sample ($R_{\HIs}$ encloses 90 per cent of each galaxy's $\HI$ mass at $z=0$). In both cases, the inflow and outflow rates have been normalized by the total gas mass of each galaxy (measured within the same aperture), and are plotted as a function of the galaxy's stellar mass. Note that, regardless of stellar mass, galaxies with asymmetric $\HI$ line profiles experience stronger gaseous outflows and more rapid gas accretion than their symmetric counterparts. Plotting conventions follow Fig.~\ref{Asymfgas}.}
    \label{flowvsmstar}
  \end{center}
\end{figure}

The results above are consistent with a picture in which feedback-driven outflows disturb the gas in low-mass galaxies and result
in asymmetric features in their $\HI$ distributions. The trend, however, is inverted at higher stellar masses, where asymmetric systems
exhibit lower than average sSFRs. Exactly why remains unclear. One possibility is that asymmetric features in the $\HI$ distributions of massive
galaxies arise as a result of disturbances driven by feedback from AGN rather than from star formation.
Another possibility is that mass growth in this regime is dominated by mergers \citep[e.g.][]{Robotham2014}, which disturb the ordered motions in galactic disks. 

Motivated by this, we examine in Fig.~\ref{flowvsmstar} the relationship between $\HI$ line asymmetry and the gas accretion and outflow rates for \eagle centrals. To do so we track gas particles between consecutive outputs, tagging those that either enter or exit a spherical aperture of radius $r=R_{\HIs}$ centered on the main progenitor of each central galaxy ($R_{\HIs}$ is the physical radius that encloses 90 per cent of each  galaxy's total $\HI$ mass at $z=0$). For the purposes of this study, we use a time interval of $\Delta t \approx 0.6$~Gyr. The specific gas outflow and inflow rates (i.e. the gas inflow and outflow rates normalized by the total gas mass of the galaxy at $z=0$ within the same aperture) are plotted as a function of stellar mass in the upper and lower panels of Fig.~\ref{flowvsmstar}, respecitvely. Note that, regardless of stellar mass, galaxies with the highest global $\HI$ line asymmetries exhibit considerably higher specific outflow rates than those with symmstric $\HI$ lines. This is true even for galaxies with stellar masses above $\approx 10^{10.3} {\rm M_\odot}$, where asymmetries in $\HI$ disks are clearly {\em not} driven by feedback from star formation (see Fig.~\ref{ssfrvsmstar}).

As mentioned above, asymmetries can also be induced by the accretion of gas, prior to its settling into a 
rotationally ordered disk. In the lower panel of Fig.~\ref{flowvsmstar} we plot the specific accretion rate of new gas particles as a function of stellar mass. As with the outflow rates, galaxies with asymmetric line profiles also exhibit elevated accretion rates, potential fuel for star formation and AGN activity. Note that the results in Fig.~\ref{flowvsmstar} are largely insensitive to the aperture size; e.g., we obtain similar trends for apertures 2 and 3 times larger. Our results therefore support the interpretation of \citet{Watts2021}, i.e. that physical processes that disturb the gas content of galaxies are the primary drivers of asymmetry. Indeed, we will see below that massive galaxies with asymmetric $\HI$ emission lines and low sSFRs typically do not harbour rotationally supported $\HI$ disks, and are often associated with early-type, dispersion-supported stellar systems.

\subsubsection{Kinematic morphology and its relation to global $\HI$ emission-line asymmetry}\label{rotsup}

Galaxies with ``disky''  morphologies are rotationally supported, suggesting a discernible link between morphology -- particularly of $\HI$
disks -- and the asymmetries of their emission-line profiles. We explore this connection below using our sample of \eagle centrals.
When calculating diagnostics for galaxy morphology, we adopt a coordinate system at rest with respect to the galaxy's centre-of-mass
motion, and align the $z$-axis
with the net angular momentum vector of the disk, $\vec{L}$ (note that $\vec{L}$ is calculated separately for $\HI$ and stars, depending
on which component is being analyzed). We denote the $z-$component of 
particle $i$'s angular momentum as $L_{z,i}$, and its distance
from the $z-$axis as $R_i=(r_i^2-z_i^2)^{1/2}$, where $r_i$ is the three-dimensional radial coordinate.

We quantify separately the morphology of $\HI$ and stars for our \eagle sample using two distinct kinematic morphology indicators.
The first, $\kappa_{\rm{ co}}$, characterises the fraction of disk's total kinetic energy, $K$, that is contributed by co-rotation \citep[e.g.][]{Correa2017}. It is defined as 
\begin{equation}
  \kappa_{\rm{co}} = \frac{1}{2\,K}\sum_{L_{z,i}>0}\, m_i\,\left(\frac{L_{z,i}}{m_i\,R_i}\right)^2,
\label{kappa}
\end{equation}
where $m_i$ is the ($\HI$ or stellar) mass of the $i^{\rm th}$ particle, and the sum is carried out over all (gas or stellar) particles with $L_{z,i}>0$ (note that the total kinetic energy of each component also includes counter rotating material).

\begin{figure}
  \begin{center}
    \includegraphics[width=1\columnwidth,trim={0cm 0cm 0cm 0cm},clip=true]{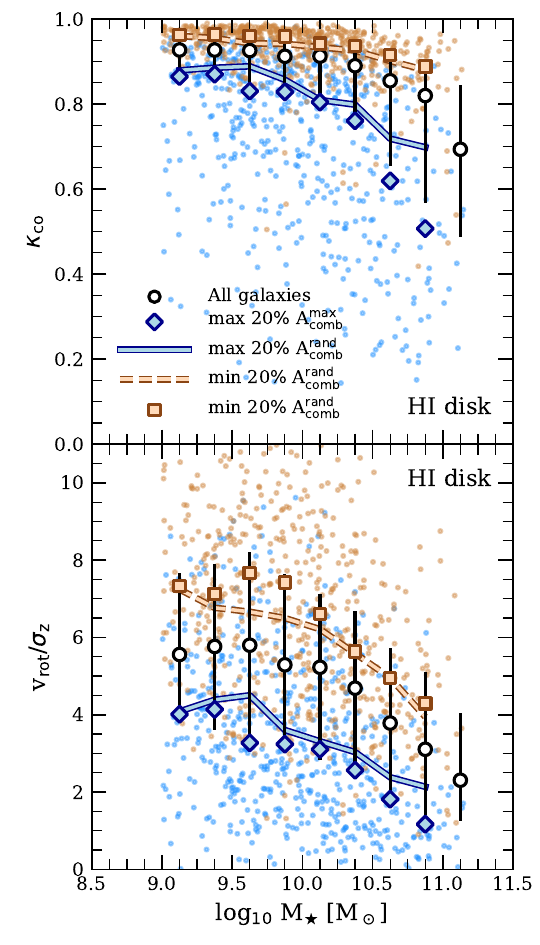}
    \caption{Two kinematic measures of the morphology of $\HI$ disks plotted versus galaxy stellar mass.
      The upper panel plots $\kappa_{\rm co}$ [equation~(\ref{kappa})], which quantifies the fraction of the $\HI$ disk's total
      kinetic energy that is contributed by co-rotating orbits; the lower panel plots the ratio of the disk's rotation velocity to its vertical velocity dispersion, i.e. $v_{\rm rot}/\sigma_z$. Plotting conventions follow Fig.~\ref{Asymfgas}.}
    \label{kappavsasym}
  \end{center}
\end{figure}

The second morphology indicator quantifies the disk's level of rotational support using the ratio of rotation-to-dispersion velocities, i.e. 
$v_\mathrm{rot}$/$\sigma_\mathrm{z}$, which are defined
\begin{equation}
  v_\mathrm{rot} = \frac{\sum_i (L_{z,i}/R_i)}{\sum_i m_i} 
\label{vrot}
\end{equation}
and 
\begin{equation}
  \sigma_z = \left[\frac{\sum_i\,m_i(v_{z,i}^2+\sigma_{{\rm T},i}^2/3)}{\sum_i \, m_i}\right]^{1/2},
\label{sig}
\end{equation}
respectively. Note that equation~(\ref{sig}) quantifies the velocity dispersion perpendicular to the plane of the disk (i.e. $v_{z,i}$ is the magnitude of the particle $i$'s velocity projected along the $z-$axis). Note also that $\sigma_{{\rm T},i}=\sqrt{k_{\rm B} T_i/m_{\rm p}}$, the three-dimensional
intrinsic thermal velocity dispersion, is not included when calculating $\sigma_z$ for stellar particles.

Fig.~\ref{kappavsasym} shows the relationship between stellar mass and $\kappa_{\rm co}$ (upper panel) and
$v_\mathrm{rot}$/$\sigma_\mathrm{z}$ (lower panel), where both quantities have been calculated for the $\HI$ disk. As in
Fig.~\ref{Asymfgas}, individual galaxies that meet our resolution criterion are shown using colored dots in both panels; the colors, symbols and line styles are chosen to match those in Fig.~\ref{Asymfgas}. 
These results indicate that the morphology of $\HI$ disks, and their degree of rotational support, correlate strongly with line
profile asymmetry. This is true for line profiles obtained for random viewing angles, as well as for lines of sight that
maximize the apparent asymmetry of $\HI$ emission line, although the strength of the correlation increases slightly for the latter. Specifically,
symmetric emission lines are associated with galaxies whose $\HI$ disks have amongst the highest values of both $\kappa_{\rm co}$ and
$v_{\rm rot}/\sigma_z$, regardless of the galaxy's stellar mass. In fact, the median value of $A_{\rm comb}^{\rm max}$ ($A_{\rm comb}^{\rm rand}$)
for systems with $\kappa_{\rm co}\geq 0.95$ is only $\approx 0.10$ (0.06), which roughly corresponds to the $20^\mathrm{st}$ ($30^\mathrm{th}$) percentile of the full distribution of asymmetries.

\begin{figure}
  \begin{center}
    \includegraphics[width=1\columnwidth,trim={0cm 0cm 0cm 0cm},clip=true]{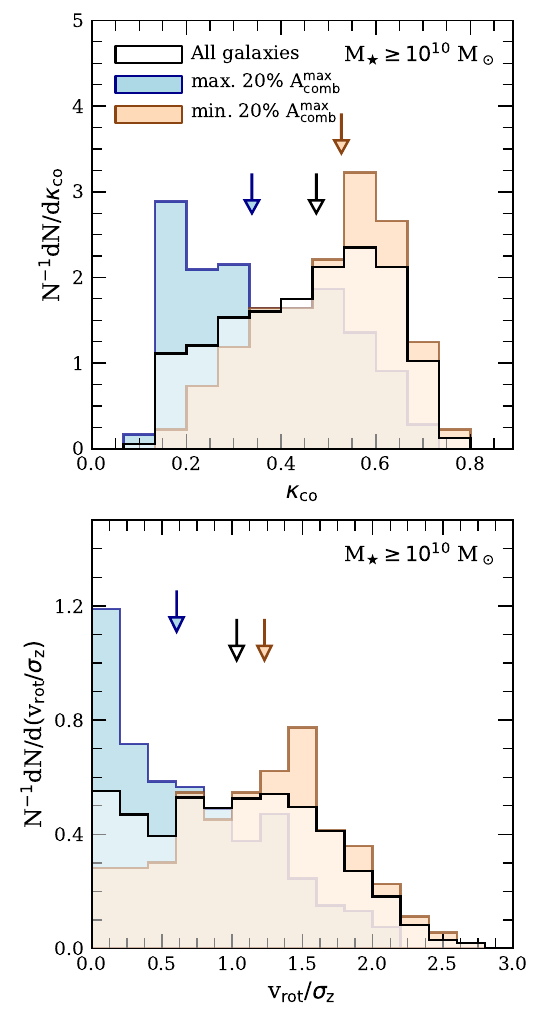}
    \caption{Probability distributions (i.e. normalized to unit area) of $\kappa_{\rm co}$
      [upper panel; equation~(\ref{kappa})] and $v_{\rm rot}/\sigma_z$ (lower panel) obtained from the stellar particles of \eagle galaxies. All galaxies are resolved
      with at least $10^3$ $\HI$-rich gas particles, and have stellar masses exceeding $10^{10}\rm{M}_\odot$ (equivalent to 
      the mass of approximately $5.5\times 10^3$ primordial gas particles). Black histograms show distribution of each quantity
      for the entire sample of \eagle galaxies that meet these criteria, whereas the blue and brown histograms are limited to the 
      subset of galaxies occupying the highest and lowest 20 per cent of $A_{\rm comb}^{\rm max}$ (note that
      the distributions obtained for extrema in $A_{\rm comb}^{\rm rand}$ are similar, but are not plotted for
      clarity). Downward pointing arrows of similar colour mark the median of each distribution. Galaxies with the symmetric $\HI$
      emission spectra typically have rotationally supported $\HI$ disks, but also
      have disk-like stellar morphologies.} 
    \label{kappavsasym_stars}
  \end{center}
\end{figure}

Interestingly, however, many of the galaxies with the {\em least} symmetric $\HI$ emission-line spectra (blue points in
Fig.~\ref{kappavsasym}) also possess
rotationally supported disks, although this is predominantly the case for the lowest-mass galaxies. For example, the
$\HI$ disks of galaxies with $M_\star\lesssim 10^{10}~\rm{M}_\odot$ that are among the sample with the least symmetric global $\HI$ lines (i.e. blue points)
have, on average, $\langle\kappa_{\rm co}\rangle\approx 0.79$ and $\langle v_{\rm rot}/\sigma_z\rangle\approx 3.89$.
High-mass galaxies with
asymmetric $\HI$ lines, however, exhibit $\HI$ morphologies consistent with dispersion-supported or turbulent structures. For example,
at $M_\star\approx 10^{11}M_\star$, galaxies with $A_{\rm comb}^{\rm max}$ below the 20$^{\rm th}$ percentile have $\langle\kappa_{\rm co}\rangle\approx 0.53$ and
$\langle v_{\rm rot}/\sigma_z\rangle\approx 1.47$, suggesting comparable levels of ordered and random motions of $\HI$ fluid elements.

In Fig.~\ref{kappavsasym_stars} we plot the distributions of $\kappa_{\rm co}$ and $v_{\rm rot}/\sigma_z$ obtained for
stellar particles. Unlike Fig.~\ref{kappavsasym}, here we restrict the sample of \eagle galaxies to those with stellar
masses exceeding $10^{10}\rm{M}_\odot$ (in addition to the resolution requirement of $\geq 10^3$ $\HI$-rich gas particles discussed in
Section~\ref{resolution}), which corresponds to the mass of approximately $5.5\times 10^3$ (primordial) gas particles. The additional restriction placed
on the stellar mass of galaxies is intended to exclude those whose stellar components are vulnerable to spurious collisional
heating, which may alter their kinematics and morphology \citep[see][for details]{Ludlow2020,Ludlow2021}. Indeed,
the results of \citet{Ludlow2019} suggests that galaxies resolved with fewer than $\approx 5000$ stellar particles are subject
to spurious collisional heating within their half-mass radii.

With these restrictions on stellar mass, the results plotted in Fig.~\ref{kappavsasym_stars} reveal a link between the morphology of a galaxy's stellar component and the shape of its $\HI$ emission-line spectrum, with symmetric line profiles most
commonly associated with galaxies whose stellar components exhibit coherent rotation and above average levels of
rotational support. The brown and blue histograms in Fig.~\ref{kappavsasym_stars}, for example, plot the distributions
of $\kappa_{\rm co}$ (upper panel) and $v_{\rm rot}/\sigma_z$ (lower panel) for the stellar components of galaxies whose
global $\HI$ emission spectra place them below and above 20$^\mathrm{th}$ and 80$^\mathrm{th}$
percentiles of $A_{\rm comb}^{\rm max}$, respectively.
Note that the medians (indicated by downward pointing arrows) differ considerably between the two samples:
$\langle\kappa_{\rm co}\rangle\approx 0.34$ ($\langle v_{\rm rot}/\sigma_z\rangle\approx 0.60$) for those with the highest
asymmetries, and $\langle\kappa_{\rm co}\rangle\approx 0.53$ ($\langle v_{\rm rot}/\sigma_z\rangle \approx 1.23$) for those with
the lowest. Whether this is consistent with the relation between $\HI$ line asymmetry 
and optical morphology in observed galaxies remains unclear \citep[see, e.g.,][]{Watts2021}.

Although the results presented above were obtained for central galaxies, we have verified that the relationship between
$\HI$/stellar morphology and the asymmetry of $\HI$ emission lines also applies to satellite galaxies in our sample. However,
given the low number of well-resolved satellites (only 200 have $M_\star\geq 10^{10}\rm{M}_\odot$ and
$\geq 10^3$ $\HI$-rich particles), we have decided not to show these results explicitly.

\subsubsection{The relationship between $\HI$ emission-line asymmetry and the dynamical state of dark matter haloes}

Galaxies are embedded within dark matter haloes, quasi-equilibrium structures that accrete mass both smoothly
and through mergers. The rate of accretion affects the dynamical state of a halo \citep{Power2012}, and mergers can
penetrate their more relaxed central regions \citep{Wang2011,Ludlow2012}, where galaxies form from gas that has cooled from the
halo's periphery. We have seen in Section~\ref{gasflows} that the asymmetries of global $\HI$ emission lines correlate strongly with the specific gas accretion rate onto galaxies. It is therefore reasonable to assume that the dynamical state of a halo may also affect the structure of the central gaseous disk, thereby influencing the asymmetry of its emission line.

\begin{figure}
  \begin{center}
    \includegraphics[width=0.95\columnwidth,trim={0cm 0cm 0cm 0cm},clip=true]{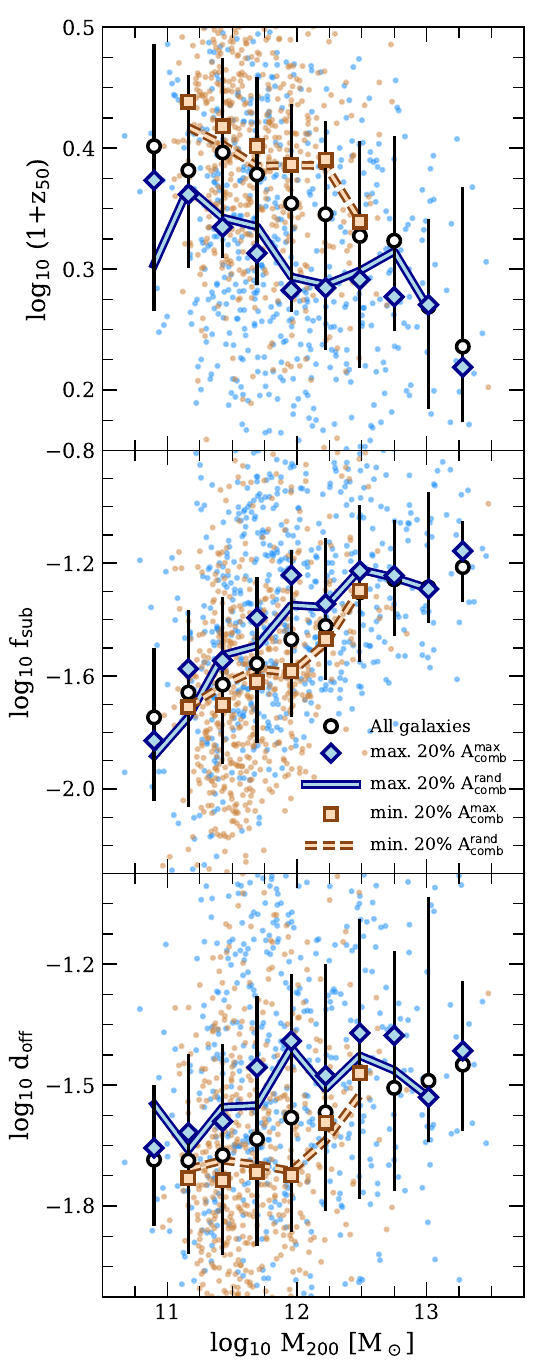}
    \caption{The half-mass formation redshift ($z_{50}$; upper panel), substructure mass fraction ($f_{\rm sub}$; middle panel)
      and centre-of-mass offset parameter ($d_{\rm off}$; lower panel) plotted as a function of of the host halo's virial mass,
      $M_{200}$, for all \eagle central galaxies resolved with at least $10^3$ $\HI$-rich gas particles and whose stellar masses exceed $10^9\,{\rm M_\odot}$. Each of these quantities serves as a proxy
      for the dynamical state of a galaxy's host dark matter halo, and therefore correlates with halo mass.
      At fixed halo mass, regardless of the LOS to a galaxy, the asymmetry of its
      $\HI$ line correlates with the formation time, substructure fraction and centre-of-mass offset of its parent dark matter
      halo in a manner that suggests that recently formed, less-relaxed halos are more likely to host asymmetric $\HI$ disks. Plotting conventions follow Fig.~\ref{Asymfgas}.}
      \label{assembly}
  \end{center}
\end{figure}

We assess the dynamical state of haloes using three quantities \citep[e.g.][]{Neto2007,Ludlow2012,Power2012}:
1) the fraction of the halo's virial mass contained in self-bound substructure, defined
\begin{equation}
  f_\mathrm{sub} \equiv \frac{\sum_{i=1}^{N_\mathrm{sub}} m_{{\rm sub},i}}{M_{200}},
\label{fsub}
\end{equation}
where $m_{{\rm sub},i}$ is the mass of the $i^\mathrm{th}$ subhalo (note that $i=0$ corresponds to the central subhalo); 2) the offset between the halo's centre of mass, $\vec{r}_\mathrm{cm}$ (calculated using all dark matter particles within $r_{200}$), and centre of potential, $\vec{r}_\mathrm{p}$, normalized to the halo's virial radius, i.e.
\begin{equation}
  d_{\rm off} \equiv \frac{|\vec{r}_\mathrm{cm}-\vec{r}_\mathrm{p}|}{r_{200}};
\label{dmoff}
\end{equation}
and 3) the half-mass formation redshift of the halo, defined by
\begin{equation}
  M_{200}(z_{50})\equiv 0.5\times M_{200}(z=0)
\label{tform}
\end{equation}
where $M_{200}(z)$ is the virial mass of the halo's main progenitor at redshift $z$.

Note that all three of these equilibrium diagnostics correlate with halo mass: more-massive haloes typically
formed more recently, contain higher levels of substructure and exhibit elevated values of $d_{\rm off}$. For that
reason, we plot in Fig.~\ref{assembly} the halo mass-dependence of these quantities, specifically highlighting
differences between haloes that host galaxies with symmetric (brown points and lines) or asymmetric $\HI$ reservoirs (blue points and lines). From top to bottom, the various panels of Fig.~\ref{assembly} plot the virial mass dependence of $z_{50}$, $f_\mathrm{sub}$, and $d_{\rm off}$, respectively. (Note that plotting conventions are inherited from Fig.~\ref{Asymfgas}.)

The results plotted in Fig.~\ref{assembly} are noteworthy for a couple of reasons. First,
they reveal that the most massive haloes tend to be occupied by galaxies with asymmetric $\HI$ line profiles. In fact,
the typical asymmetry for galaxies hosted by haloes with masses $\gtrsim 10^{12.5}\,{\rm M_\odot}$ is about $50$ per cent larger than
for galaxies occupying halos with $M_{200}\lesssim 10^{12}\,{\rm M_\odot}$. Second, regardless of halo mass, galaxies with the highest
asymmetries tend to occupy significantly less-relaxed haloes with more recent formation times and higher levels of
substructure than those hosting galaxies with symmetric $\HI$ emission lines. It is therefore likely that the shapes of $\HI$ emission
lines are sensitive not only to baryonic processes, but also reflect potential gravitational disturbances due to, for example,
the recent accretion of dark matter, or gravitational perturbations due to elevated levels of substructure. 

\begin{figure}
  \begin{center}
    \includegraphics[width=1\columnwidth,trim={0cm 0cm 0cm 0cm},clip=true]{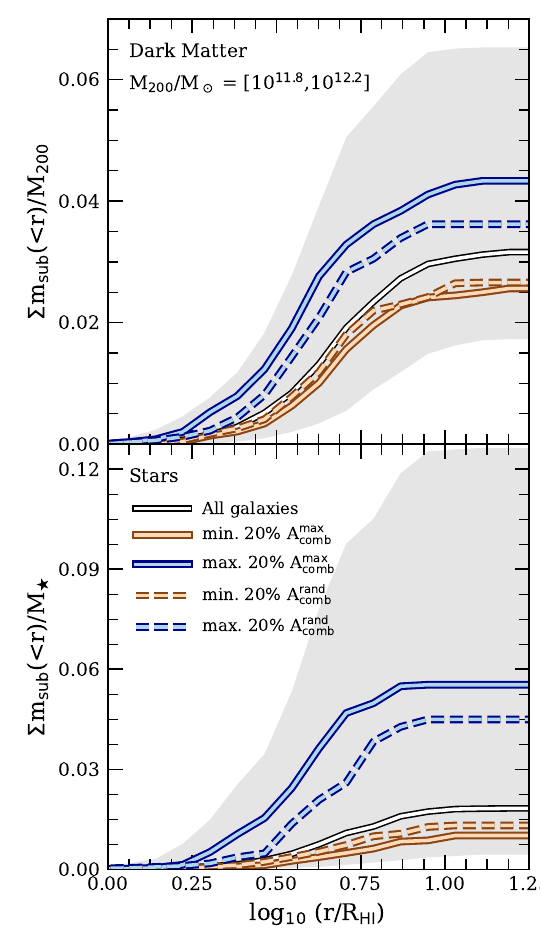}
    \caption{Radial profiles of the cumulative mass fraction in self-bound substructure surrounding central galaxies whose host dark matter haloes span the mass range $11.8\leq \log_{10} (M_{200}/{\rm M_\odot})\leq 12.2$. The integrated mass profiles are shown separately for dark matter substructures (upper panel; here the mass is expressed in units of $M_{200}$), and for the self-bound stellar mass in satellite galaxies (lower panel; in this case, masses are normalized to the stellar mass of the central galaxy). The white curve is the median profile for all centrals in the quoted mass range, and the shaded region is $20^\mathrm{th}$--$80^\mathrm{th}$ percentile halo-to-halo scatter. Solid blue and solid brown curves are the median profiles for the galaxies in upper and lower 20 per cent of $A_\mathrm{comb}^\mathrm{max}$, for quoted mass bin, respectively. Dashed blue and dashed brown curves are similar to the solid curves, but for galaxies in the 20 per cent extremes of $A_\mathrm{comb}^\mathrm{rand}$. In all cases, radii have been normalized by $R_{\HIs}$, i.e. the radius that encloses 90 per cent of the galaxy's total $\HI$ mass. Note that, at any given radius, asymmetric centrals are surrounded by considerably more substructure in the form of dark matter subhaloes and satellite galaxies than their symmetric counterparts.}
    \label{subprof}
  \end{center}
\end{figure}

In Fig.~\ref{subprof} we show that the higher levels of substructure surrounding asymmetric galaxies are not limited to the outer regions of the halo, nor are the substructure fractions dominated by subhaloes unable to form stars due to their low mass. Here we plot radial profiles of the cumulative substructure mass fraction for dark matter subhaloes (upper panel) and luminous satellites (lower panel), where the profile is normalized by $M_{200}$ for the former, and by the stellar mass of the central galaxy for the latter. Because of the strong mass-dependence of $f_{\rm sub}$, we focus on central galaxies whose hosts span a narrow mass range, $11.8\leq \log_{10} (M_{200}/{\rm M_\odot})\leq 12.2$, but find similar trends for other halo mass bins. In all cases, radii have been normalized by $R_{\HIs}$. The results indicate that haloes hosting asymmetric centrals have significantly more mass contributed by dark matter substructure (by about a factor of 1.8) and luminous satellites (by about a factor of 5) within their virial radii, although there are measurable differences \textit{within all radii}. The differences are noticeable even within a few characteristic disk scale lengths, $R_{\HIs}$.

\subsection{The asymmetries of the $\HI$ line profiles of satellite galaxies}\label{satellites}

Until now, we have focussed the majority of our analysis on the emission line asymmetries of well-resolved central galaxies
in \eagleNS. However, as highlighted in Fig.~\ref{cenvsatasym}, the global asymmetries of central and satellites differ
slightly but systematically; satellites are, on average, less symmetric than centrals (see also \citealt{Watts2020b}).
This is not unexpected; the latter are exposed to a variety of physical processes that
are unlikely relevant for centrals, including tidal or ram pressure stripping \citep[e.g.][]{Marasco2016}, and
close encounters with nearby satellites \citep{Bosch2017,Bakels2021}. 

What impact do these physical processes have on the asymmetries of the $\HI$ line profiles of satellite galaxies? To shed light on this, we next examine which, if any, of these environmental processes impact satellite asymmetries. We adopt an approach similar to that employed by \citet{Marasco2016} for estimating their importance, each of which we now describe.

\begin{enumerate}
\item {\em Ram pressure stripping:} We estimate the relative importance of ram pressure to that of the gravitational restoring force acting on the $\HI$ disks of satellite galaxies using following \citet{Gunn1972}. Specifically, disks are vulnerable to ram pressure stripping if
  \begin{equation}
    \rho_{\rm IGM}\, v_{\rm rel}^2>\Sigma_{\rm ISM}(R)\,\left.\frac{\partial\Phi(R,z)}{\partial z}\right\vert_{z=0},
    \label{eq:ram}
  \end{equation}
  where $\rho_{\rm IGM}$ is the density of the IGM at the satellite's location, $v_{\rm rel}$ is its relative velocity   with respect to the IGM, $\Phi(R,z)$ is the total gravitational potential at the point $(R,z)$ (i.e. at a radius $R$ within the disk's midplane, and a height $z$ above it) within the satellite,   $\partial\Phi/\partial z\vert_{z=0}$ is the corresponding gravitational acceleration directed towards the disk midplane, and $\Sigma_{\rm ISM}(R)$ is the ISM 
  surface density at $R$. The ratio between the left- ($P_{\rm ram}$) and right-hand ($P_{\rm grav}$) sides of equation~(\ref{eq:ram}) can be used to assess whether $\HI$ disks may be influenced by ram pressure forces.

  We estimate $P_{\rm ram}$ using the mass-weighted density ($\rho_{\rm IGM}$) and relative velocity ($v_{\rm rel}$) of the 500 gas particles that are nearest to the satellites centre of mass and also bound to the satellite's host dark matter halo (note that we exclude gas particles
  that are bound to the satellite itself, or to other nearby satellites). As discussed by \cite{Marasco2016}, this provides a useful estimate of the density and bulk motion of the IGM in the satellite's vicinity (typically on scales of tens of kpc to $\approx 100\,{\rm kpc}$), while avoiding the possible contamination due to gas particles bound to nearby satellites
  (satellite--satellite interactions are considered below).

\begin{figure*}
  \begin{center}
    \includegraphics[width=2.1\columnwidth,trim={0cm 0cm 0cm 0cm},clip=true]{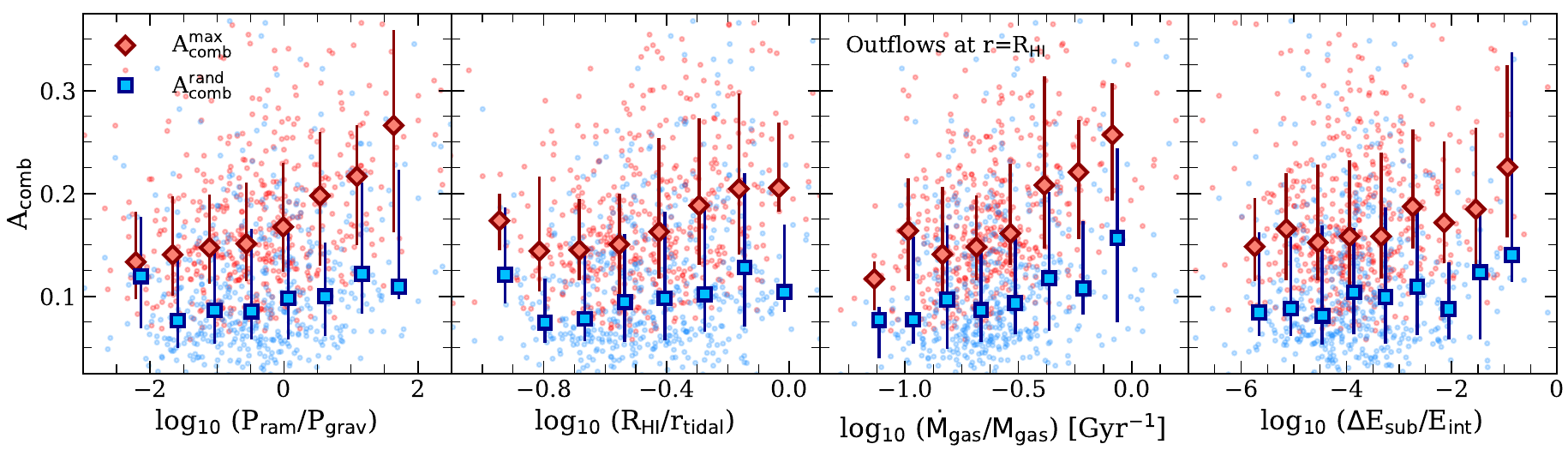}
    \caption{From left to right, different panels show the impact of ram pressure (equation~\ref{eq:ram}), tidal stripping (equation~\ref{eq:rtide}), gas outflows, and satellite--satellite encounters (equation~\ref{eq:delE}) on the combined asymmetry parameter, $A_{\rm comb}$ (equation~\ref{eq:Acomb}). Results are plotted for satellite galaxies in the \eagle simulation that are resolved with $\geq 10^3$ $\HI$-rich gas particles. Each of these quantities are proxies for environmental effects and are described in detail in Section~\ref{satellites}. Faint blue and red dots in each panel correspond to $A_{\rm comb}^{\rm rand}$ and $A_{\rm comb}^{\rm max}$, respectively, measured for individual satellite galaxies; their medians and 20$^\mathrm{th}$--80$^\mathrm{th}$ percentile scatter are shown using connected symbols and error bars of corresponding colour. Note that global $\HI$ line profiles are systematically more asymmetric among satellite galaxies affected by ram pressure and tidal stripping, but are much less sensitive to energy injection due to  satellite--satellite interactions. The effects are most pronounced when quantified in terms of the  maximal asymmetry, $A_{\rm comb}^{\rm max}$.}
    \label{stripvsasym}
  \end{center}
\end{figure*}

  We calculate the ISM surface density and mid-plane gravitational restoring force for each satellite galaxy at the radius $R_{\HIs}$ that encloses 90 per cent of its $\HI$ mass. This is comparable to the radius at which the $\HI$ column density reaches $\approx 1 \,{\rm M_\odot \, pc^{-2}}$, where most hydrogen is in atomic form; the local ISM surface density at $R_{\HIs}$ is therefore
  $\Sigma_{\rm ISM}(R_{\HIs})\approx \Sigma_{\HIs}(R_{\HIs})/X_\mathrm{H}\simeq 1/X_\mathrm{H}\, {\rm M_\odot \, pc^{-2}}$ 
  ($X_\mathrm{H}=1-Y\simeq 0.752$ is the hydrogen fraction). To estimate the mid-plane gravitational restoring force, we first orient each satellite galaxy so that the $z$-axis is coincident with the minor axis of the inertia tensor of all stellar particles with $R_{\HIs}$. We then calculate the vertical gravitational acceleration directed towards the mid-plane at 36 points, evenly-spaced in azimuthal angle and at a radius $R=R_{\HIs}$. To do so, we calculate the total gravitational potential at $z=0$ (i.e. in the disk mid-plane) and at a height $z=2\epsilon$ above the disk, where $\epsilon=2.66\,{\rm kpc}$ is the (Plummer-equivalent) gravitational softening length (note that all particles bound to the satellite -- dark matter and baryonic -- are included in the calculation of the gravitational potential). The net gravitational acceleration toward the mid-plane -- shown as the potential gradient on the right-hand side of equation~(\ref{eq:ram}) -- is then approximated as $[\langle\Phi(R_{\HIs},2\,\epsilon)\rangle-\langle\Phi(R_{\HIs},0)\rangle]/(2\,\epsilon)$, where the chevrons indicate averages over all 36 points.

\item {\em Tidal stripping:} In a typical gas-rich galaxy, the $\HI$ disk extends considerably farther than the stars,
  and is therefore more vulnerable to tidal stripping. Because features in $\HI$ line profiles often arise as a result of
  bulk motions in the outer disk, it is plausible that tidal stripping may reduce their asymmetries by preferentially
  stripping disturbed gas at large galacto-centric radii. Alternatively, tides may deform symmetric disks thereby increasing the
  asymmetries of their emission-line profiles.

  We estimate the tidal radii of all satellite galaxies using \citep[see e.g.][]{Tormen1998}
  \begin{equation}
    r_\mathrm{tidal} = r_{\rm sub}\Biggl[\frac{m_{\rm sub}}{(2-\,{\rm d}\ln M/{\rm d}\ln r|_{r \rightarrow r_{\rm sub}})\,M(r_{\rm sub})}\Biggr]^{1/3},
    \label{eq:rtide}
  \end{equation}
  where $r_{\rm sub}$ is the satellite's radial separation from its host, $m_{\rm sub}$ is its total mass, and $M(r_{\rm sub})$ is the total enclosed mass of the host within $r_{\rm sub}$. Equation~(\ref{eq:rtide}) provides a useful approximation for the radius beyond which the satellite's mass is likely to be stripped due to tidal forces exerted by its host halo \citep[see e.g.][]{Springel2008}.

  As an additional proxy for the combined effects of ram pressure and tidal stripping, we also measure the outflow rates of gas particles through the surface of a sphere of radius $r=R_{\HIs}$ centred on the main progenitor of each satellite galaxy (how the
  outflows are calculated was previously described in Section~\ref{gasflows} for central galaxies; we use the same procedure for satellites). Although additional processes (e.g. AGN or stellar feedback) may contribute to outflows, we do not attempt to distinguish their physical origin.

\item {\em Satellite--satellite encounters:} Encounters between satellites in groups and clusters are common
  \citep{Tormen1998,Bosch2017,Bakels2021} and the associated tidal shocks
  can result in impulsive heating. An approximate treatment of the effect follows from the distant-tide
  approximation \citep{Spitzer1958}, which predicts a change in a satellite's specific internal energy due to
  the encounter of
  \begin{equation}
    \Delta E_\mathrm{sub} \approx \frac{4\, G^2\, {\rm M_p}^2\, m_{\rm sub}}{3\, v_{\rm p,rel}^2\, b^4} \langle r^2\rangle.
    \label{eq:delE}
  \end{equation}
  Here ${\rm M_p}$ is the mass of the perturber, $v_{\rm p,rel}$ is its relative velocity with respect to the satellite, $b$
  is the impact parameter of the interaction, and the quantity $\langle r^2\rangle=\sum_i m_i\, r_i^2/\sum_i m_i$ is the mass-weighted mean-square radius of particles bound to the satellite. The derivation of equation~(\ref{eq:delE}) assumes that $b\gg r_s$ ($r_s$ being the characteristic size of the satellite) and that $v_{\rm p,rel}$ is constant, neither of which are strictly valid during genuine satellite--satellite encounters. Nevertheless, \citet{Marasco2016} showed that equation~(\ref{eq:delE}) provides a useful proxy for the ongoing effect of satellite interactions and that satellites with the highest $\Delta E_{\rm sub}$ often exhibit visually disturbed $\HI$ distributions. We follow their lead, and use the ratio of $\Delta E_{\rm sub}$ to the satellite's total internal (kinetic plus potential) binding energy, $E_{\rm int}$,  to quantify the possible importance of ongoing encounters between satellites of the same host halo. In practice, we approximate $b$ as the current separation between the satellite and the perturber in question, and for each satellite calculate $\Delta E_{\rm sub}$ due to all other resolved subhaloes of the same host. We take the maximum value of $\Delta E_{\rm sub}$ as our proxy for impulsive heating (this is reasonable because a single perturber tends to dominate the integrated sum of $\Delta E_{\rm sub}$ due to all possible satellite interactions; \citealt{Marasco2016}).
\end{enumerate}

In Fig.~\ref{stripvsasym}, we plot the dependence of the combined asymmetry parameter, $A_{\rm comb}$, on proxies for each of the environmental processes mentioned above. From left to right, the various panels focus on $P_{\rm ram}/P_{\rm grav}$, $r_{\rm tidal}$ (expressed, for convenience, as the ratio $R_{\HIs}/r_{\rm tidal}$), gaseous outflows, and the ratio $\Delta E_{\rm sub}/E_{\rm int}$. In all panels, individual subhaloes are shown as coloured points: red and blue distinguish $A_{\rm comb}^{\rm max}$ from $A_{\rm comb}^{\rm rand}$, respectively; the corresponding median trends are indicated using connected diamonds and squares of similar colour, respectively (error bars highlight the 20$^\mathrm{th}$ to 80$^\mathrm{th}$ percentile scatter). For comparison, we also plot using dashed lines the median relations for the 127 satellites whose host haloes have masses that exceed $M_{200}=10^{13}\,\rm{M}_\odot$.

The results plotted in Fig.~\ref{stripvsasym} suggest that ram pressure and tidal stripping both contribute
to the global line profile asymmetries of satellite galaxies, with ram pressure perhaps being the dominant effect;
impulsive heating, however, is negligible, at least for the range of satellite interactions probed by our sample.
For example, we find that $A_{\rm comb}^{\rm max}$ (which is not biased by projection effects) increases systematically with
increasing $P_{\rm ram}/P_{\rm grav}$, with evidence of a steepening trend for $P_{\rm ram}/P_{\rm grav}\gtrsim 1$,
albeit with increasing scatter as well (on average $\langle A_{\rm comb}^{\rm max}\rangle \approx 0.22$ for
$P_{\rm ram}/P_{\rm grav}\geq 1$, whereas $\langle A_{\rm comb}^{\rm max}\rangle \approx 0.17$ for $P_{\rm ram}/P_{\rm grav} < 1$).
The Spearman rank correlation\footnote{The Spearman rank test quantifies the strength of the correlation between two quantities while being agnostic to the order of the correlation, i.e. it tests for both linear and non-linear trends. All the Spearman rank coefficients mentioned in this paper correspond to a $p$-value of $<0.003$ for the null hypothesis of no correlation, which corresponds to a $>3\sigma$ confidence level.} coefficient is 0.33, which is somewhat larger than for $A_{\rm comb}^{\rm rand}$ (0.24).

Although the correlations in the second panel of Fig.~\ref{stripvsasym} 
are weak and exhibit considerable scatter (the Spearman rank
coefficient is 0.31 for $A_{\rm comb}^{\rm max}$; 0.20 for $A_{\rm comb}^{\rm rand}$), there is a clear increase
in the typical asymmetries of line profiles among systems whose tidal radii are comparable to $R_{\HIs}$. For example,
satellites for which $R_{\HIs}/r_{\rm tidal}\gtrsim 0.8$ have average maximal asymmetries of
$\langle A_{\rm comb}^{\rm max}\rangle\approx 0.24$, whereas those with $R_{\HIs}/r_{\rm tidal}\lesssim 0.2$ have
$\langle A_{\rm comb}^{\rm max}\rangle\approx 0.15$ ($\langle A_{\rm comb}^{\rm rand}\rangle$ is $\approx 0.14$ and
$\approx 0.10$ for the same two samples, respectively).

Note, however, that ram pressure and tidal stripping act in unison, and their 
combined effects should also be apparent in gas flow rates. Indeed, the second panel from  right in
Fig.~\ref{stripvsasym} shows that the $\HI$ line asymmetry correlates 
more strongly with specific gas outflow rates than with either ram pressure or tidal stripping alone
(note that outflow rates for satellites are calculated as described in Section~\ref{gasflows} for centrals). For example, the Spearman rank correlation coefficient
is $0.41$ for $A_{\rm comb}^{\rm max}$, and $0.23$ for $A_{\rm comb}^{\rm rand}$.

The connection between environment and visual disturbances in the gas content of satellites has also been discussed
in previous studies based on hydrodynamical simulations. For example, \citet{Marasco2016} used the \eagle simulation
to assess the impact of tidal and ram pressure stripping, and satellite encounters on visual disturbances in the
$\HI$ content of satellite galaxies. They found that ram pressure due to the IGM of the surrounding host dark matter
halo is the dominant process that disturbs a satellite's $\HI$, which is in line with our results above. 
Similarly, \citet{Yun2019} visually identified satellites with disturbed gas reservoirs in IllustrisTNG and found that, on 
average, they experience stronger ram pressure forces than the general satellite population. They reported
that satellites of more massive haloes are more prone to ram pressure stripping, as also shown by 
\citet{Stevens2019} for IllustrisTNG. 

\section{Summary}\label{summary}
The high incidence of galaxies observed to exhibit asymmetric $\HI$ emission lines demands a thorough investigation of their physical origins. In this work we addressed this issue using the \eagle simulation \citep{Schaye2015,Crain2015}. We generated mock 21-cm emission-line profiles for well-resolved galaxies with $\HI$ and stellar masses similar to those of central galaxies in the xGASS survey \citep{Catinella2018}. We verified that our mock line profiles are physically realistic by comparing their asymmetries to those of observed emission lines in xGASS, finding good agreement. We then investigated the origins of their asymmetric features by exploring the sensitivity of our mock $\HI$ emission-line profiles to observational effects, and how asymmetry is related to various drivers of galaxy evolution.

Our main results can be summarised as follows:
\begin{enumerate}

\item Gas particles associated with \eagle galaxies typically exhibit a bimodal distribution of $\HI$ fractions ($f_{\HIs}$), with populations of particles that are either $\HI$-deficient or $\HI$-rich (Fig.~\ref{fhihist}). More than 90 per cent of the $\HI$ mass of a typical \eagle galaxy is locked-up in $\HI$-rich particles, which we define to be those with $\HI$ fractions $f_{\HIs}\geq 0.5$. The asymmetries of global $\HI$ line profiles can be measured robustly (i.e. with sampling-induced random errors $\lesssim 22$ per cent) provided galaxies are resolved with $\gtrsim 10^3$ $\HI$-rich particles (see Fig.~\ref{pois}). We therefore only included galaxies resolved with $\geq 10^3$ $\HI$-rich particles in our analysis. 

\item The inferred asymmetries of unresolved $\HI$ lines are sensitive to the line's effective velocity resolution ($\Delta v_{\rm{sm}}$), the amount of instrumental noise ($\sigma_{\rm{rms}}$) and the distance from the source to the observer ($D$; see Section~\ref{edgeasym}). Larger values of either of these parameters reduces the lines profile's signal-to-noise ratio ($S/N$), resulting in a higher inferred asymmetry (see Fig.~\ref{obillust} for one example). Asymmetry is typically more sensitive to instrumental noise and distance than to velocity resolution. For example, modest changes to $\sigma_{\rm{rms}}$ that are compatible with the intrinsic
  noise variation among xGASS observations (spanning, for example, the $25^{\rm th}$ to $75^{\rm th}$ percentile in $\sigma_{\rm rms}$) can affect the inferred asymmetry by more than a factor of $\approx 4$ (middle panel of
  Fig.~\ref{obillust}). However, when imposing values of $\Delta v_{\rm{sm}}$ and $\sigma_{\rm{rms}}$ that are compatible with
  modal values of xGASS observations (and assuming sensible galaxy distances), we find that the line profiles of xGASS-like samples of \eagle galaxies -- i.e. those that match the stellar and $\HI$ mass distribution of galaxies in the xGASS survey -- exhibit distributions of $\HI$ line asymmetries that are consistent with those inferred from xGASS observations (see Fig.~\ref{xgasymcomp}). 

\item In agreement with \citet{Deg2020}, we find that global $\HI$ line asymmetries are sensitive to projection effects.
  Mock line profiles constructed assuming low inclinations (i.e. for face-on, or nearly face-on
  projections of $\HI$ disks) often exhibit large asymmetries. However, such profiles are not resolved with a sufficient number of velocity channels
  to properly sample the velocity structure of the $\HI$ disk, and the resulting line profiles are dominated by the (Gaussian) thermal component
  of gas particle velocities. As a result, their asymmetry estimates are unreliable and do not reflect the underlying spatial or kinematic
  distribution of $\HI$ in the galaxy's disk. When low inclinations (e.g. $\lesssim 40$ degrees) are avoided,
  asymmetries typically attain maximum values when galaxies are viewed edge-on, and we exploit this feature to obtain an estimate of the
  {\em maximum} intrinsic asymmetry of the $\HI$ line profiles of all \eagle galaxies in our sample. We find that maximum asymmetries obtained for edge-on
  orientations  ($A_\mathrm{comb}^\mathrm{max}$) are on average a factor of $\approx 1.9$ higher than those inferred for random sight lines
  ($A_{\rm comb}^{\rm rand}$; Fig.~\ref{cenvsatasym}). 

\item The $\HI$ line asymmetries of \eagle galaxies are largely independent of their stellar mass, regardless of viewing angle
  (Fig.~\ref{AsymMstar}). This confirms the observational result of \citet{Watts2020} based on galaxies in the xGASS
  survey. However, galaxies with asymmetric $\HI$ emission lines generally have lower $\HI$ fractions than their symmetric
  counterparts; but they have higher ${\rm H_2}$ fractions (Fig.~\ref{Asymfgas}), at least for stellar masses $\lesssim 10^{10.3}$~M$_\odot$. Because the fraction of ${\rm H_2}$ in gas particles (Section~\ref{himod}) correlates strongly with their local density and metallicity -- two quantities that also dictate star formation in \eagle -- we also find considerable differences in the star formation rates (SFRs) of symmetric and asymmetric galaxies. Specifically, below
  $M_\star\approx 10^{10.3}$~M$_\odot$, galaxies with asymmetric line profiles exhibit considerably higher SFRs than
  average. At higher stellar masses, however, the trend reverses: asymmetric galaxies above $M_\star\approx 10^{10.3}$~M$_\odot$
  have lower $\HI$ and ${\rm H_2}$ fractions, and lower star formation rates (see Fig.~\ref{ssfrvsmstar}). Exactly why
  remains unclear, but we speculate that below $\approx 10^{10.3}$~M$_\odot$ line asymmetries of \eagle galaxies reflect disturbances
  in gaseous disks driven by stellar feedback, and at higher masses by AGN feedback or mergers. 

\item Support for this interpretation is provided by stark differences in gas outflow rates measured in the regions
  surrounding symmetric and asymmetric galaxies; in fact, the latter experience outflow rates that exceed the former by a factor
  of $\approx 2$, regardless of stellar mass. Galaxies with asymmetric $\HI$ disks also experience substantially
  greater gas {\em inflow} rates than those with symmetric $\HI$ disks (see Fig.~\ref{flowvsmstar}). It is therefore
  unsurprising that asymmetry correlates strongly with various indicators of the morphology of $\HI$ disks. Specifically,
  symmetric galaxies have considerably higher levels of ordered rotation [as quantified by the kinematic morphology
  parameter $\kappa_{\rm co}$; equation~(\ref{kappa})], and also higher rotation-to-dispersion velocity ratios (Fig.~\ref{kappavsasym}).
  Interestingly, we also find similar correlations between the asymmetry of $\HI$ lines and the morphology of the {\em stellar}
  component of galaxies: those with high asymmetry are more likely to be associated with dispersion-supported stellar
  systems; those with low asymmetry to galaxies with disky morphologies (Fig.~\ref{kappavsasym_stars}).

\item The asymmetry of $\HI$ emission lines is also sensitive to the dynamical state of a galaxy's surrounding
  dark matter halo. For example, regardless of halo mass, galaxies with asymmetric $\HI$ lines typically reside in unrelaxed
  halos with more recent formation times and higher levels of substructure (Fig.~\ref{assembly}). Asymmetric centrals are also
  associated with a larger number of luminous satellite galaxies (Fig.~\ref{subprof}).

\item In agreement with \citet{Watts2020b} (see also \citealt{Watts2020}), we find that satellite galaxies in \eagle typically have
  more asymmetric $\HI$ line profiles than centrals (see Fig.~\ref{cenvsatasym}), likely due to the different environmental
  processes that affect their evolution. In order to test this, we quantified three separate environmental processes -- ram pressure
  stripping due to the surrounding IGM, tidal stripping due to the background host halo, and dynamical heating due to satellite--satellite
  encounters -- and assessed their impact of on the asymmetries of the $\HI$ lines of satellites. We find that ram pressure and
  tidal stripping are strong contributors to satellite asymmetry. Encounters between satellites are not (Fig.~\ref{stripvsasym}). 

\end{enumerate} 

Though our results shed light on the origin of $\HI$ emission-line asymmetries, there are important caveats that warrant
consideration. First, the $\HI$ and ${\rm H_2}$ content of \eagle galaxies was determined in post-processing using a simple
empirical model that does not include a self-consistent treatment of the cold and warm phases of the ISM. Although the predictions of the model agree well with observations in some respects \citep{Bahe2016,Crain2017}, the model neglects several important processes, such as ionization due to local sources and cosmic rays, the formation of ${\rm H_2}$ on dust grains and its destruction due to Lyman-Werner photons. We note, however, that other prescriptions -- such as the ones proposed by \citet{K13} and \citet{GD14}, in which the impact of Lyman-Werner photons from local sources can be included -- give similar results to the ones presented in this paper.

\eagle is among the few simulations with sufficient (mass and spatial) resolution and volume to carry out studies such as
ours, but it nevertheless offers limited statistics: only a few thousand galaxies are sufficiently well-resolved to allow their $\HI$
line profiles to be modelled robustly, most of them centrals. This precludes binning galaxies in multiple dimensions which, in our opinion, is essential if we are to disentangle the various physical processes that give rise to the diversity of line profile asymmetries. For example, the results presented in Figs~\ref{Asymfgas} to \ref{assembly} indicate that there is considerable overlap in any one galaxy or halo property and the asymmetry of its line profile. This suggests that there are likely multiple drivers of asymmetry, and unravelling them will likely require a larger-volume simulation than \eagle but of comparable (or better) mass resolution.

These limitations prohibited us from carrying out ``strong convergence'' tests of $\HI$ line asymmetries, and \eagleNS's mass resolution restricted our analysis to galaxies that are relatively gas-rich. This, combined with the limited volume of \eagleNS,
places constraints on the diversity of galaxies used in our study and the environments in which they form. There is thus considerable
scope for future studies of line profile asymmetries using larger-volume and/or higher-resolution simulations that can overcome the
these limitations, which may provide a more holistic picture of the origins of $\HI$ line asymmetry.

\section*{ACKNOWLEDGEMENTS}
We thank Luca Cortese and Adam Watts for helpful discussions, and Claudia Lagos for providing merger trees for \eagle galaxies. We would also like to thank Dr. Nathan Deg for a useful referee report. AM and RW acknowledge support from the Australian Government through Research Training Program (RTP) Scholarships. ADL and ASGR are supported by the Australian Research Council through the Future Fellowship scheme (project IDs: FT160100250 and FT200100375, respectively). ARHS is the recipient of the Jim Buckee Fellowship at The University of Western Australia. This work has utilized computational resources provided by Pawsey Supercomputing Centre\footnote{\url{https://pawsey.org.au/}} with funding from the Australian Government and the Government of Western Australia. The plots in this paper are produced using
the \textsc{\large matplotlib} package for \textsc{\large python} \citep{Hunter2007}. Other key \textsc{\large python} packages
used in this work are \textsc{\large numpy} \citep{Numpy}, \textsc{\large scipy} \citep{Scipy} and \textsc{\large astropy}
\citep{Astropy}. The paper has been typeset using the online freemium academic writing environment
Overleaf\footnote{\url{https://www.overleaf.com/}}. 

\section*{Data Availability}
The observational results presented in the paper are based on publicly available data obtained from the xGASS survey webpage (\url{http://xgass.icrar.org/data.html}).
Simulation results are based on publicly available data obtained from the \eagle project website (\url{http://icc.dur.ac.uk/Eagle/database.php}).
The \textsc{\large python} routines that were used for generating and analyzing the $\HI$ lines
are available at \url{https://github.com/adimanuwal/GAHILE}. The gas accretion and outflow flow rates 
were calculated using the code that is available at \url{https://github.com/RJWright25/hydroflow}.

\bibliographystyle{mnras}
\bibliography{citations}

\appendix
\section{Correlations between the various asymmetry measurements}\label{asymcor}
\begin{figure*}
    \begin{center}
        \includegraphics[width=2.1\columnwidth,trim={0.0cm 0.0cm 0cm 0.0cm},clip=true]{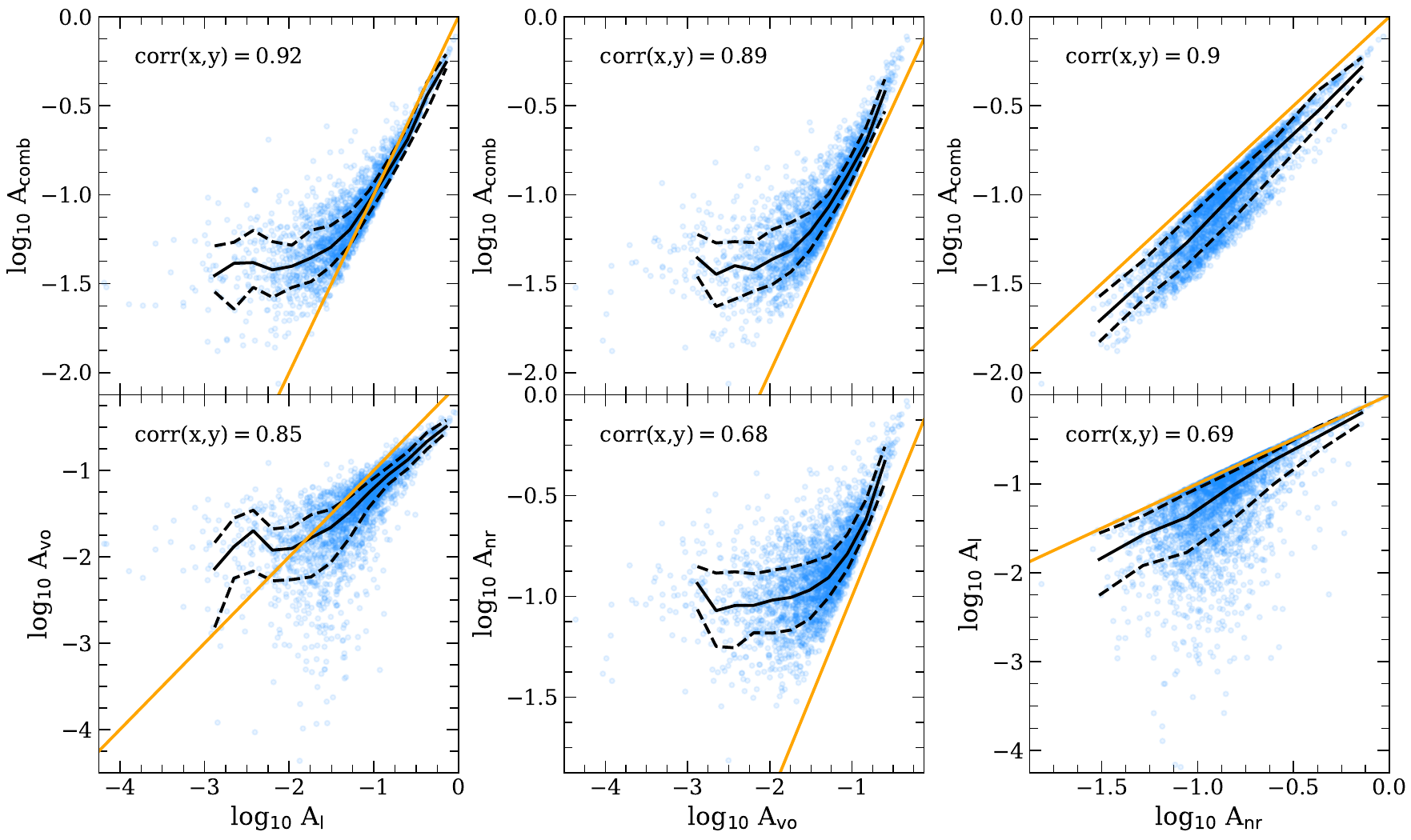}
    \caption{Correlations between $A_\mathrm{l}$, $A_\mathrm{vo}$, $A_\mathrm{nr}$ and $A_\mathrm{comb}$ for random orientations of central galaxies in \eagle. The top row shows the correlations between $A_\mathrm{comb}$ and all other statistics described in Section~\ref{measure}. The bottom row shows the correlations between $A_\mathrm{l}$, $A_\mathrm{vo}$, $A_\mathrm{nr}$. In each panel, the solid black curve shows the medians, and the dashed black curve shows the $20^{\rm th}$ and $80^{\rm th}$ percentiles. The solid orange line shows the 1:1 relation. The Spearman rank correlation coefficients are provided in the
    top-left corner of each panel.}
    \label{asymcorrs}
    \end{center}
\end{figure*}
In Fig.~\ref{asymcorrs} we plot the correlations between the various asymmetry statistics (described in Section~\ref{measure}) obtained for the $\HI$ profiles of central galaxies in \eagle (note that for a particular galaxy, all asymmetry statistics were obtained for the same random LOS). The Spearman rank correlation coefficients are shown in the various panels. It is clear that all asymmetry statistics correlate strongly with one other. Note, however, that the scatter between them typically increases at low asymmetries, and decreases at high asymmetries. This implies that an $\HI$ line that is asymmetric in any one statistic is likely to be asymmetric in the other statistics, but this is not the case if it is instead very symmetric in one of them. 
This is because each asymmetry measure carries a different information about line asymmetry: $A_{\rm nr}$, for example, is sensitive to both large- and small-scale variations in the line profile that may arise as a result of, e.g., $\HI$ substructure in the disk, or instrumental noise; $A_{\rm vo}$ and $A_{\rm l}$ on the other hand are based on integrated fluxes on the low- and high-velocity sided of some central velocity, and are therefore more sensitive to global asymmetries in line profile than local ones.

The bottom row of Fig.~\ref{asymcorrs} shows that the $A_\mathrm{vo}$--$A_\mathrm{l}$ correlation is the strongest, which is not surprising given that $A_\mathrm{vo}$ is connected to $A_\mathrm{l}$ by construction: larger differences in the integrated fluxes between the two sides of $v_\mathrm{sys}$ imply larger offsets between $v_\mathrm{sys}$ and $v_\mathrm{eq}$. The $A_\mathrm{nr}$--$A_\mathrm{vo}$ and $A_\mathrm{l}$--$A_\mathrm{nr}$ relations have similar strengths, both have Spearman rank coefficients of $\approx 0.7$. The top-row of Fig.~\ref{asymcorrs} shows that $A_\mathrm{comb}$ is strongly correlated with all three statistics.

Note also that $A_{\rm nr}$ is larger than both $A_{\rm vo}$ and $A_{\rm l}$, which implies $A_{\rm comb}\leq A_{\rm nr}$. The fact that $A_\mathrm{l}\leq A_\mathrm{nr}$ can be trivially understood. The denominators in equation~(\ref{eq:Al}) (which defines $A_{\rm l}$) and equation~(\ref{eq:nres}) (which defines $A_{\rm nr})$ are equivalent, and the numerator in equation~(\ref{eq:Al}) can be written as
\begin{multline}
\left|\sum_{v=v_{\rm l}}^{v_{\rm sys}}\,F(v)\Delta v - \sum_{v=v_{\rm sys}}^{v_{\rm h}}\,F(v)\Delta v\right|\\
= \left|\sum_i [F(v_{\rm sys}-i\Delta v) - F(v_{\rm sys}+i\Delta v)]\Delta v\right|.
\end{multline}
It is well known that for real numbers $a$ and $b$, that $|a+b|\leq|a|+|b|$. This implies
\begin{multline}
\left|\sum_i [F(v_{\rm sys}-i\Delta v) - F(v_{\rm sys}+i\Delta v)]\Delta v\right|\\
\leq \sum_i \left|F(v_{\rm sys}-i\Delta v) - F(v_{\rm sys}+i\Delta v)\right|\Delta v.
\label{identity}
\end{multline}
The right hand side of equation~(\ref{identity}) is the numerator in equation~(\ref{eq:nres}), and the left-hand side the numerator in equation~(\ref{eq:Al}), which implies $A_\mathrm{l}\leq A_\mathrm{nr}$.

\bsp	
\label{lastpage}
\end{document}